\documentclass[prd,twocolumn,showpacs,preprintnumbers,amsmath,amssymb,nofootinbib]{revtex4}
\usepackage{stmaryrd}
\usepackage{multibox}
\usepackage{graphicx, epic, eepic, color}
\usepackage{dcolumn}
\usepackage{bm}
  \renewcommand{\a}{\alpha}
  \renewcommand{\b}{\beta}
  
  \renewcommand{\d}{\delta}
    
    \newcommand{\e}{\epsilon}
    
    \newcommand{\h}{\eta}
    \newcommand{\G}{\Gamma}
  \renewcommand{\l}{\lambda}

  \renewcommand{\k}{\kappa}
  
    \newcommand{\p}{\varphi}
  
    \newcommand{\s}{\sigma}
  
   \newcommand{\ve}{\varepsilon} 
    \newcommand{\x}{\xi}

\def\pa{\partial}

   \newcommand{\RR}{\mathbb{R}}

    \def\pr{\partial}

    \newcommand \om[1]{\textmd{${}^{\textmd{\tiny (#1)}}\omega$}{}}
    \newcommand \omhat[1]{\textmd{${}^{\textmd{\tiny (#1)}} \hat
    \omega$}{}}
    \newcommand \omhathat[1]{\textmd{${}^{\textmd{\tiny
    (#1)}} \hat{\hat \omega}$}{}}
     \newcommand \Ga[1]{\textmd{${}^{\textmd{\tiny (#1)}}\Gamma$}{}}
    \newcommand \Gahat[1]{\textmd{${}^{\textmd{\tiny (#1)}} \hat \Gamma$}}

    \definecolor{green}{rgb}{0.0,0.7,0.0} 
    \definecolor{grey}{rgb}{1.,0.9,0.95} 

    \def\p1{\phantom{1}}
    \def\IR{{\hbox{{\rm I}\kern-.2em\hbox{\rm R}}}}

\begin{document}

\preprint{HUTP-06/A0004}

\title{A Higgs Mechanism for Gravity. Part II: Higher Spin Connections}

\author{Nicolas Boulanger}
\affiliation{\\ Universit\'e de Mons-Hainaut, Acad\'emie Wallonie-Bruxelles,\\
M\'ecanique et Gravitation, Avenue du Champ de Mars 6, B-7000 Mons, Belgium}

\author{Ingo Kirsch}
\affiliation{\\ Jefferson Laboratory of Physics, Harvard University,\\  
Cambridge, MA 02138, USA }

\begin{abstract} 
We continue the work of hep-th/0503024 in which gravity is considered
as the Goldstone realization of a spontaneously broken diffeomorphism
group. We complete the discussion of the coset space
\textit{Diff}$(d,\RR)/SO(1,d-1)$ formed by the $d$-dimensional group
of analytic diffeomorphisms and the Lorentz group. We find that this
coset space is parameterized by coordinates, a metric and an infinite
tower of higher-spin-like or generalized connections.  We then
study effective actions for the corresponding symmetry
breaking which gives mass to the higher spin connections. Our model
predicts that gravity is modified at high energies by the exchange of
massive higher spin particles.
\end{abstract}


\maketitle

\section{Introduction}

Gauge field theories have a long and successful history in elementary
particle physics.  As it is generally known, the starting point for
gauging is the experimental observation of a conserved charge which,
via Noether's theorem, is related to a rigid symmetry. In particular,
a conserved energy-momentum current corresponds to the invariance
under global space-time translations. Since energy-momentum is the
source of gravity, one expects the gravitational interaction to emerge
from gauging the global translational symmetry. Indeed, general
relativity (GR) can be derived by gauging the translational group as
was first conclusively shown in \cite{Cho:1975dh}. The gauge status of
gravity remains however rather subtle, see e.g.\
\cite{Fronsdal:1989ci, MAG, Blagojevic:2002du, Tresguerres:2002uh,
Sardanashvily:2005vd} and many references therein.

In this paper we adopt the view that gauge theories of gravity
describe only the low-energy effective, {\textit{i.e.}} massless degrees 
of freedom of a more general gravitational theory featuring a
spontaneously broken space-time symmetry. This view is based on a
theorem in~\cite{I&O} which states that the gauge theory associated
with a local group $G_{loc}$ can be obtained by the nonlinear
realization of the corresponding infinite-parameter group $G$ with the
Poincar\'e group $H$ being the vacuum stability group. 
The Minkowski metric $\h_{ij}$ is assumed to be given from the beginning. 
Mechanisms for the selection of the signature of the metric have
been proposed in \cite{Itin,Leclerc}. 

Let us illustrate the theorem for the case in which $G_{loc}$ is the
local translational group $T(d)$ in $d$ dimensions. By construction,
the group $T(d)$ is locally isomorphic to the infinite-parameter group
$\textit{Diff}\,(d,\RR)$ of analytic diffeomorphisms
\cite{Ogievetsky}.  Then, according to the theorem of \cite{I&O}, the
(simplest) gauge theory of the translational group $G_{loc}=T(d)$,
general relativity, can be derived by nonlinearly realizing
$G=\textit{Diff}\,(d,\RR)$. 

The key observation in the proof of this theorem is that the gauge
potential of $G_{loc}$, here the tetrad $e_{i}^{~j}$, can be identified
with a parameter of the coset space $G/H$. In fact, applying the
nonlinear realization method \cite{CCWZ} adapted to space-time groups
\cite{Salam,Isham:1971dv,Volkov,Ogie}, one finds that $G/H$ is
parametrized by the field $h^i_{~j}$ and an infinite tower of fields
$\om{s}^i{}_{j_1...j_sk}$ \mbox{($ s \geqslant 1$)}.  The exponential
of $h^i_{~j}$, $e^{~i}_{j} \equiv (e^h)^i_{~j}$, transforms
exactly as a tetrad \cite{Isham:1971dv}. The translational gauge
potential thus arises as one of the Goldstone fields of a
spontaneously broken diffeomorphism invariance.

This was first explicitly shown by Borisov and Ogievetsky
\cite{Borisov}.  These authors used the fact that the
infinite-dimensional algebra of analytic diffeomorphisms can be
represented as the closure of two finite-dimensional algebras
\cite{Ogievetsky}.  This splitting, however, impeded the discussion of
the remaining Goldstone fields $\om{s}$. It was pointed out later in
\cite{I&O} that the fields $\om{s}$ may acquire mass through a Higgs
effect, leaving the tetrads as the only massless degrees of freedom.
Since gauging translations only provides the tetrads but not the
massive fields $\om{s}$, the gauge principle leads to the correct
low-energy effective theory, at least as long as the masses of
$\om{s}$ are high enough. However, if one takes the idea of a
spontaneously broken diffeomorphism invariance seriously, the
gravitational interaction will be modified at high energies by the
exchange of massive fields $\om{s}$.

This paper is devoted to the study of these coset fields and the
construction of a spontaneous symmetry breaking mechanism \cite{Higgs}
for the diffeomorphism group. (For brevity, we will refer to it as
``Higgs mechanism'', although this terminology is unfair, see
Refs.~\cite{Higgs}.) To gain some insight into this Higgs mechanism,
we first complete the nonlinear realization of \textit{Diff}$(d,\RR)$
studied in \cite{Borisov,Pashnev,Kirsch} by providing the complete
transformation laws for all coset fields of
\textit{Diff}$(d,\RR)/SO(1,d-1)$.  We find that the fields $\om{s}$
naturally generalize the concept of a linear connection; the
transformation law is inhomogeneous and contains the $s+1$th
derivative of the diffeomorphism parameter $\ve^i(x)$.

The nonlinear realization will also provide gauge transformations of
the (linearized) form
\begin{align}
\om{1}'{}_{ij k} &= \om{1}_{i j k} - \pr_k  h_{ij} \,, \label{a1}\\
\om{s}'{}^i{}_{j_1...j_s k} &= \om{s}^i{}_{j_1...j_s k} - 
 \pr_k  \om{s-1}^i{}_{j_1...j_s} \quad(s \geqslant 2) \,, \label{a2}
\end{align}
which show a mutual absorption of the Goldstone fields
$\om{s}^i{}_{j_1...j_s k}$ and $h_{ij}$. The generalized connection
$\om{s}$ of level $s$ eats the connection $\om{s-1}$ of level $s-1$,
while the ordinary connection $\om{1}^i{}_{jk}$ absorbs the metric
$h_{ij}$. This will give mass to each of the generalized connections
$\om{s}$ ($s \geqslant 2 $) and $\om{1}_{(ij)k}$. As the field with
the lowest spin in the coset space, the metric remains massless.

In the second part of the paper we model the gravitational Higgs
mechanism by concrete actions. We restrict here to find actions for
the lowest two absorption processes given by Eqs.~(\ref{a1}) and
(\ref{a2}) for $s=1,2$. We consider these models as describing only a
part of the full Higgs mechanism for the complete diffeomorphism group
which, in full generality, appears to be quite complex.

The first model we present describes the breaking of the linear
group $GL(d,\RR) \subset \textit{Diff}\,(d,\RR)$ down to the Lorentz
group $SO(1,d-1)$. Here we assume that the generalized connections
$\om{s}$ with $s \geqslant 2$ have already been decoupled and we are
left with a massless linear connection $\om{1}^i{}_{jk}$ of an
effective affine space-time.  The breaking of the tangential group
$GL(d,\RR)$ will then be induced by the introduction of the metric as
a Higgs field. This involves the absorption process (\ref{a1}) by
which the symmetric part $\om{1}'_{(ij)k}$ of the connection acquires
mass.

The model is largely based on that given in \cite{Kirsch}. There are
two essential improvements: i) We explicitly show that the Higgs
mechanism leads to a massive spin-3 field associated with the totally
symmetric field $\om{1}'_{(i j k)}$. ii)~The field which was
introduced in \cite{Kirsch} as a kind of gravitational analog to the
(so-called) Higgs particle plays now the role of an auxiliary field in the
Singh-Hagen formulation \cite{Singh:1974qz} of the massive spin-3
field. We therefore do not predict a new Higgs particle.

The second model describes the absorption process (\ref{a2}) for $s=2$
by means of which the field $\om{2}'{}^i{}_{j_1j_2k}$ becomes massive.
The model is along the lines of the so-called ``telescopic Higgs
effect'' (see \cite{Aulakh} and references therein), more recently also
known as ``La Grande Bouffe" \cite{Bianchi}. Here we aim at the more
modest goal of constructing the St\"uckelberg Lagrangian for the massive
field $\om{2}'{}^i{}_{j_1j_2k}$.

In the last part of the paper, we discuss a possible relation between
the coset fields $\om{s}$ and higher spin connections as first introduced    
in the gauge formalism of higher spin fields in \cite{Vasiliev:1980as}. 
Note that a relation between the latter formalism and the nonlinear 
realization approach was recently pointed out in \cite{Vasiliev:2005zu}.
We also show that a space endowed with generalized connections
satisfies the strong equivalence principle and is equivalent to a
space-time with a velocity-dependent affine connection.

The paper is organized as follows. In section~\ref{sec2} we study the
coset space \textit{Diff}$(d,\RR)/SO(1,d-1)$ by means of the nonlinear
realization approach \cite{CCWZ}. We also discuss the double role of
Goldstone fields in gravity as absorber fields and fields which get
absorbed by other Goldstone fields. In section~\ref{sec3} we construct
Higgs models which lead to a ultraviolet modification of general
relativity. In section~\ref{sec4} we discuss a possible link between the
generalized connections $\om{s}$ and higher spin connections known from
the literature. We also discuss the geometrical structure of a
space-time equipped with generalized connections. We conclude in
section~\ref{sec5} with some final remarks and open questions.
%
%
\section{Nonlinear realization of the analytic diffeomorphism group} 
\label{sec2}
%
In this section we consider the (left) coset space
\textit{Diff}$(d,\RR)/SO(1, d-1)$ formed by the $d$-dimensional group
of analytic diffeomorphisms \textit{Diff}$(d,\RR)$ and its stabilizing
Lorentz subgroup $SO(1, d-1)$. We show that the coset space is
parametrized by a coordinate field, a metric and an infinite tower of
generalized connections.

\subsection{Review on the diffeomorphism algebra}

We begin by briefly reviewing the algebra of analytic diffeomorphisms.
The diffeomorphism algebra is generated by an infinite tower of
generators $F^{(m)}_i{}^{j_1...j_{m+1}}$ ($m=-1,...,\infty$) which are
symmetric in the $m+1$ upper indices. The lowest generators are the
translations $P_i\equiv F_i^{(-1)}$ and the generators of the linear
group $L_i{}^j\equiv F^{(0)}_i{}^j$. Generators
$F^{(m)}_i{}^{j_1...j_{m+1}}$ with $m \geqslant 1$ generate nonlinear
transformations.

The corresponding diffeomorphism algebra \textit{diff}$(d,\RR)$ is given
by the commutation relations
\begin{align} \label{alg}
&[F^{(n)}_k{}^{i_1...i_{n+1}}, F^{(m)}_l{}^{j_1...j_{m+1}}] = \nonumber \\
   &\hspace{.5cm}= i \sum\limits_{a=1}^{m+1} \delta^{j_a}_k 
      F^{(m+n)}_l{}^{i_1...i_{n+1}j_1...{\hat j}_a...j_{m+1} } \nonumber\\
   &\hspace{.5cm}- i \sum\limits_{a=1}^{n+1} \delta^{i_a}_l 
      F^{(m+n)}_k{}^{i_1...\hat i_a...i_{n+1}j_1...j_{m+1} }  ,
\end{align}
where indices with a hat are omitted. We easily identify the
Lorentz (sub-)algebra
 \begin{align}
[M_{ij}, M_{kl}]&=i\eta_{j[k} M_{l]i}- i\eta_{i[k}M_{j]l} \,,
\end{align}
with Lorentz generators $M_{ij}\equiv L_{[i}{}^k \eta_{j]k}=
F^{(0)}_{[i}{}^k \eta_{j]k}$. We denote complete strength-one
antisymmetrization on indices by using square brackets, while complete
strength one symmetrization is denoted by curved brackets.  For
example, $F^{(0)}_{[i}{}^k \eta_{j]k}\equiv \frac{1}{2}
(F^{(0)}_{i}{}^k \eta_{jk}-F^{(0)}_{j}{}^k \eta_{ik})\,$.

\subsection{The coset space \textit{Diff}$(d,\RR)/SO(1, d-1)$
and generalized connections} \label{secIIb}

The coset space $G/H=\textit{Diff}\,(d,\RR)/SO(1, d-1)$ is
parametrized by the fields $\xi^i$ ($d$ parameters), $h_{ij}$
($d(d+1)/2$ parameters) and an infinite set of fields
$\om{s}^i{}_{j_1...j_{s}k}$ \mbox{$(s \geqslant 1)$}, each with
\begin{align} 
  d \begin{pmatrix} d+s \\s+1 \end{pmatrix}
\end{align} 
components. These fields are associated with broken translations
$P_i$, shear transformations and dilations $T_{ij}= L_{(ij)}$, and
generators $F^{(s)}_i{}^{j_1...j_{s}k}$ $(s \geqslant 1)$,
respectively.

As was shown in \cite{Pashnev, Kirsch} the parameters $\xi^i$
transform as coordinates under the diffeomorphism group,
\begin{align}\label{diffinv}
\delta \xi^i = \varepsilon^i(\xi) \,,
\end{align}
with $\varepsilon^i(\xi)$ the parameters of \textit{Diff}$(d,\RR)$.
As explained in detail in \cite{Kirsch}, the breaking of translations
makes the parameters of \textit{Diff}$(d,\RR)$ dependent on the
coordinates $\xi^i$ and turns the coset parameters into space-time
dependent fields. For a recent discussion on the tight link between
the coset fields $\xi^{i}$ and space-time coordinates, see
\cite{Tresguerres:2002uh}.

The transformation behavior of the coset field $h_{ij}(\xi)$ has been
known since the very first publications on nonlinear realizations of
space-time groups \cite{Salam, Borisov}.  It is usually given for the
exponential
\begin{align}
  e^{~i}_{j} \equiv (e^h)^i{}_j =
  \d^i_j + h^i{}_j + h^i{}_k h^k{}_j/2 + \ldots
\end{align}
which transforms as a tetrad \cite{Borisov}. {}From now on we will be
using Greek indices for the Minkowski metric $\eta_{\alpha\beta}$ and
define a space-time metric, as usual, by
\begin{align} \label{metric}
  g_{ij} = e_i{}^\alpha e_j{}^\beta \eta_{\alpha\beta} \,.
\end{align}

Finally, the field $\omega^i{}_{jk}(\xi)$ associated with
$F^{(1)}_i{}^{jk}$ was shown to transform as a linear connection under
the diffeomorphism group \cite{Pashnev, Kirsch},
\begin{align} \label{trafoomega}
  \delta \omega^i{}_{j k}=\,& \frac{\partial \ve^i}{\partial \xi^m}
  \omega^m{}_{j k} -  2 \frac{\partial \ve^m}{\partial \xi^{(j}}
  \omega^i{}_{k)m}
  + \frac 1 2 \frac{\partial^2 \ve^i}{\partial \xi^{j}\partial \xi^{k}} \,.
\end{align}
Since $\omega^i{}_{jk}$ is symmetric in the indices $j$ and $k$,
there is no torsion.


So far not much attention has been paid to the fields
$\om{s}^i{}_{j_1...j_{s}k}$ associated with the nonlinear generators
$F^{(s)}$ with $s>1$.  These fields are completely symmetric in the
lower $s+1$ indices, which is ultimately a consequence of the assumed
commutativity of the coordinates of $\RR^d$~\cite{Kirsch}.

Using the general nonlinear realization technique
\cite{CCWZ,Salam,Isham:1971dv,Volkov,Ogie,Borisov}, in App.~\ref{app1}
we compute the infinitesimal transformation law for the fields
$\om{s}^i{}_{j_1...j_{s}k}$. We obtain
\begin{align}
  \delta_\ve \! \om{s}^i{}_{j_1 \ldots j_{s}k}=\,& \ve^i{}_m 
  \om{s}^m{}_{j_1 \ldots j_{s} k}
  - (s+1)  \ve^m{}_{(j_1} \om{s}^i{}_{j_2\ldots j_{s}k)m}
  \nonumber\\
  &+\ve^i{}_{j_1 \ldots j_{s}k} + {\cal O}(\om{s\textmd{-}1}) \,, 
  \label{transformation}
\end{align}
with
\begin{align}
  \ve^i{}_{j_1 \ldots j_s}=\frac{1}{s!}
  \frac{\partial^s \ve^i}{\partial{\xi^{j_1}}
  \cdots  \partial{\xi^{j_s}}} \,,
\end{align}
which generalizes Eq.~(\ref{trafoomega}) ($s=1$) to arbitrary values
of~$s$.  The l.h.s.\ of Eq.~(\ref{transformation}) is defined to be
$\delta_\ve \! \om{s}^i{}_{j_1 \ldots j_{s}k}= {\om{s}'}^i{}_{j_1
\ldots j_{s}k}(\xi')-\om{s}^i{}_{j_1 \ldots j_{s}k}(\xi)$.  We
identify Eq.~(\ref{transformation}) with the transformation behavior
of a {\em generalized connection}: The first line in
Eq.~(\ref{transformation}) is the tensor part of the transformation,
while the first term in the second line shows the inhomogeneity which
contains the $s+1$th derivative of the diffeomorphism parameter
$\varepsilon^i(\xi)$~\cite{Pashnev}.  The finite form of the
transformation law is given by
\begin{align} \label{finitetrafo} 
&\om{s}'{}^{i}_{j_1...j_sk} 
=
\frac{\partial \xi'{}^i}{\partial \xi^{m}}
\frac{\partial \xi^{l_1}}{\partial \xi'{}^{j_1}}
\cdots
\frac{\partial \xi^{l_s}}{\partial \xi'{}^{j_s}}
\frac{\partial \xi^{n}}{\partial \xi'{}^{k}}
\om{s}^{m}_{l_1...l_sn} \\
&\,\,\,\,\,\,-
\frac{\partial \xi^{l_1}}{\partial \xi'{}^{j_1}}
\cdots
\frac{\partial \xi^{l_s}}{\partial \xi'{}^{j_s}}
\frac{\partial \xi^{n}}{\partial \xi'{}^{k}}
\frac{\partial^{s+1} \xi'{}^i}{\partial \xi^{l_1} \cdots \partial \xi^{l_s} 
\partial \xi^n}
 + {\cal O}(\om{s\textmd{-}1})\,. \nonumber
\end{align}
Upon substituting $\xi'{}^i =\xi^i+\varepsilon^i(\xi)$ with
$\varepsilon^i(\xi)$ small into (\ref{finitetrafo}) and redefining
$\om{s} \rightarrow - (s+1)! \om{s}$, we regain the infinitesimal
transformation~(\ref{transformation}).

A new feature of the generalized connections is the occurrence of
additional terms in the transformation law~(\ref{transformation})
which are summarized in ${\cal O}(\om{s\textmd{-}1})$. 
By using a convenient bracket notation, in App.~\ref{app1} 
we give an algorithm to compute the complete
transformation $\d \om{s}$ including all terms in ${\cal
O}(\om{s\textmd{-}1})$. In general, these terms contain connections of
lower spin. For instance, the transformation law for the connection
$\d \om{2}^i{}_{j_1j_2k}$, Eq.~(\ref{dom2}), contains the term
\begin{align}
  2\ve^i{}_{l(j_1} \omega^l{}_{j_2k)} \,.
\end{align}
This term involves the ordinary linear connection $\omega^i{}_{j k}$
which has one index less than $\om{2}^i{}_{j_1j_2k}$.  Generalized
connections mix and cannot be considered independently from each
other.

In App.~\ref{app3} we decompose the fields $\om{s}^i{}_{j_1...j_{s}k}$
with respect to the general linear group and determine their spin
content.  We find that, unless further constraints are imposed, these
fields describe several states of different spin, where the highest
state possesses spin $s+2$. For instance, the highest component of a
general linear connection $\omega^i{}_{jk}$ ($s=1$) has spin 3, see
e.g.\ \cite{MAG}.\footnote{For a general linear connection
$\omega^i_{~jk}$, the totally symmetric component $\omega_{(ijk)}$ is
non-vanishing.  See \cite{MAG3} for recent works in metric-affine
theory of gravity, where exact solutions are built that display a
propagating spin-3 component of the linear connection. }  This leads
us to the presumption that the generalized connections $\om{s}$ are
related to higher spin connections. For the construction of actions,
we will implicitly assume this relation.  We will come back to the
possible link with higher spin connections in Sec.~\ref{secIVa}.

\begin{table}[ht]
\begin{center}
\begin{tabular}{cccccc}
  \hline {\bf broken symmetry} &{\bf generators}& {\bf geometrical field} \\ 
   \hline \smallskip
 translations & $P_i \equiv F^{(-1)}_i$ 
&coordinates $\xi^i$\\
shears/dilations &$T_{ij} \equiv F^{(0)}_{(ij)}$ &
   metric $g_{ij}$\\ 
   nonlinear  & $F^{(s)}_i{}^{j_1...j_sk}$ $(s \geqslant 1)$&
   gen.\ connections  \\ 
   transformations &  & $\om{s}^i{}_{j_1...j_sk}$\\
  \hline
\end{tabular}
\end{center} \caption{Goldstone fields parameterizing the infinite-dimensional
coset space $\textit{Diff}\,(d, \RR)/SO(1,d-1)$. } \label{tab}
\end{table}

In Tab.\ \ref{tab} we summarize the parameters of the coset space
$G/H=\textit{Diff}\,(d, \RR)/SO(1,d-1)$ and give their geometrical
interpretation. We have shown that the fields $\xi^i$, $h_{ij}$,
$\om{1}^i{}_{jk}$, $\om{2}^i{}_{jkl}$, etc.\ can be regarded as
coordinates, metric, and an infinite tower of generalized connections,
respectively. As we will see in Sec.~\ref{sec3}, most of these
Goldstone bosons become massive though and decouple at low
energies. Einstein gravity results as the appropriate low-energy
effective theory of gravity.

As in \cite{Kirsch} we define the spacetime manifold $\cal M$ as that
part of the coset space $G/H$ which is spanned by the (global)
translations.  This part is parameterized by the coordinates $\xi^i$.
If we wish to recover Einstein gravity at low energies, we need to
have {\em local} Poincar\'e invariance in the tangent space of the
manifold $\cal M$. Local translational invariance is ensured by the
diffeomorphism invariance of the manifold $\cal M$, cf.\
Eq.~(\ref{diffinv}).\footnote{$G$ can be considered as a principal
$H$-bundle over $G/H$, $\pi:G \rightarrow G/H$. Recall from
\cite{Kirsch}, Sec.~IIIA that we gain local translational invariance
on $\cal M$ (i.e.\ on the base space $G/H$) at the expense of loosing
global translational invariance in the fiber.}  Local Lorentz
invariance is more subtle to see. Note that the vacuum stability group
$H \subset \textit{Diff}\,(d, \RR)$ is just the {\em global} Lorentz
group. However, in the present nonlinear realization the group $H$
induces {\em local} Lorentz transformations: Recall that in the
transformation law for coset elements $\sigma \in G/H$ \cite{CCWZ,
Salam},
\begin{align}
g \sigma (\xi) = \sigma (\xi') h(\xi, g) \,,
\end{align}
the elements $h \in H$ depend nonlinearly on $g \in G$ and the coset
parameters $\xi$. Since global translations are broken, the group
elements $h$ depend in particular on the coordinates $\xi^i$, ie.\
they are functions of $\xi^i$, $h=h(\xi^i,...)$. We can thus perform
an independent Lorentz transformation at each spacetime point.

\subsection{The total nonlinear connection}

Let us now turn to the total nonlinear connection one-form $\Gamma$
which can be expanded in the generators of $G=$\textit{Diff}$(d,\RR)$
as
\begin{align} \label{expansion}
  \Gamma=i \vartheta^\a P_\a + i\sum_{s=1}^\infty 
  \Ga{s}^\alpha{}_{\b_1...\b_s} F^{(s-1)}_\alpha{}^{\b_1...\b_s} \,.
\end{align}
In order to find the coefficients $\vartheta^\a$ and
$\Gamma^\a{}_{\b_1...\b_{s}}$ $(s \geqslant 1)$, we fix the stabilizing
group to be $H=SO(1,d-1)$ such that an element $\sigma$ of the coset
space $G/H$ is parametrized by
\begin{align} 
  \sigma =\,&  e^{i \x^m P_m } e^{i h^{ij} T_{ij}}  
  e^{i \omega^i{}_{j_1j_{2}} F^{(1)}_i{}^{j_1j_{2}} } 
  \cdots\nonumber \\ 
  &\times e^{i \om{s}^i{}_{j_1...j_{s+1}} 
  F^{(s)}_i{}^{j_1...j_{s+1}} } \cdots \,. \label{cosetelement}
\end{align}

The coefficients of the total nonlinear connection $\Gamma
=\sigma^{-1}d \sigma$ are then given by the one-forms
\begin{align}
  \vartheta^\alpha&=(e^{-1})_k{}^\alpha  d\xi^k \,, \label{coeff-1}\\
  \Ga{1}^\alpha{}_{\beta} &= \,(e^{-1})_k{}^{\alpha}d e^k{}_\beta   
  - 2 \, \omega^\alpha{}_{\beta \gamma} \vartheta^\gamma \,, \label{coeff-2}\\
  \Ga{2}^\alpha{}_{\beta_1\beta_{2}} &= \,d \omega^\alpha{}_{\beta_1\beta_2}  
  - 3 \, \omega^\alpha{}_{\beta_1\beta_2 \gamma} \vartheta^\gamma
  - \omega^\d{}_{\beta_1\beta_2} \omega^\alpha{}_{\d\gamma}\vartheta^\gamma \nonumber\\
  &\,\,\,\,\,+2\omega^\alpha{}_{\d(\beta_1}
  \omega^\d{}_{\beta_2)\gamma}  \vartheta^\gamma
   + (e^{-1})_l^{~\alpha}d e^l{}_\gamma 
  \omega^\gamma{}_{\beta_1\beta_2}   \,,\nonumber\\
  &\,\,\,\,\,- 2 \omega^\alpha{}_{\gamma(\beta_1|} 
  (e^{-1})_l{}^{\gamma}d e^l{}_{|\beta_2)} 
\end{align}
and for $s \geqslant 3$ by
\begin{align}
  \Ga{s}^\alpha{}_{\beta_1...\beta_{s}} &=
  d \om{s\textmd{-}1}^\alpha{}_{\beta_1...\beta_s}  
  - (s+1) \, \om{s}^\alpha{}_{\beta_1...\beta_s \gamma} \vartheta^\gamma  \label{redef} 
  \nonumber\\
  &\,\,\,\,\, + {\cal O}(\om{s\textmd{-}1}{}^2)\,.
\end{align}
Here we used Latin ($i,j,...$) and Greek ($\alpha, \beta, ...$)
letters for holonomic and anholonomic (frame) indices, respectively.
${\cal O}(\om{s\textmd{-}1}{}^2)$ denotes terms of quadratic order and
higher in $\om{s\textmd{-}1}$, $\om{s\textmd{-}2}$, ..., $\om{1}$.  A
general formula for the coefficients
$\Ga{s}^\alpha{}_{\beta_1...\beta_{s}}$ is given to all orders by
Eq.~(\ref{SchoutenGamma}) in App.~\ref{appB}. To evaluate
Eq.~(\ref{SchoutenGamma}), one may use the bracket notation
introduced in App.~\ref{app1}.

As spelled out in \cite{Kirsch}, the coefficients $\vartheta^\alpha$
and $\Ga{1}^\alpha{}_{\beta}$ can be interpreted as the coframe and
linear connection. Linear connections can also be obtained by the
gauging of the linear group as first proposed in \cite{Yang} and
elaborated on in Metric-Affine Gravity \cite{MAG}, see also
\cite{Jackiw}. We observe that nonlinear realizations of
\textit{Diff}$(d,\RR)$ provide an alternative derivation of the linear
connection. 

Note that it is not possible in the nonlinear realization approach to
single out a single generalized connection (or a finite number of such
connections). Assume we break only a single generator $F^{(s)}$ which
gives rise to a single connection $\om{s}$. Then, terms of higher
order would be absent in Eqs.~(\ref{transformation}) and
(\ref{redef}). However, the stabilizing subgroup $H$ is not closed in
this case, since e.g.\ the commutator $[F^{(s-1)}, F^{(1)}]$ ends on
$F^{(s)}$. The nonlinear generators $F^{(s)}$ ($s\geqslant 1$) can
thus only be broken as a whole. This property is shared by the higher
spin algebras, see e.g.\ \cite{Fradkin:1987ks} and refs.\ therein.

\subsection{Higgs phenomenon and double role of Goldstone fields in
gravity} \label{secIId}

For the following it is useful to recall the Higgs phenomenon in
elementary particle physics. For instance, in $U(1)$ gauge theory the
gauge boson $A_\mu$ (spin 1) becomes massive due to the absorption of
a Goldstone scalar $\phi$. Usually this is achieved by the $U(1)$
gauge transformation
\begin{align}
  A'_\mu = A_\mu + \partial_\mu \phi \, \label{absorptionA}
\end{align}
turning the Goldstone field $\phi$ into the longitudinal mode of the
massive gauge boson $A'_\mu$.  

In the coset realization under consideration, the gravitational analog
of Eq.~(\ref{absorptionA}) is given by the coefficients of the total
nonlinear connection $\Gamma$. Eqs.\ (\ref{coeff-2})--(\ref{redef})
can be regarded as redefinitions of the generalized
connections. There are basically two
absorption processes:
\begin{itemize}
\item[(I)] \mbox{$s=1$}: The ordinary spin connection $\om{1}^\alpha{}_{\beta
  k}$ absorbs the degrees of freedom of the tetrad $e^i{}_\alpha$ as
  can be seen from Eq.~(\ref{coeff-2}). Since the tetrads are related
  to the shear and dilation parameters, this corresponds to the
  breaking of $GL(d,\RR)$ down to $SO(1,d-1)$. 

\item[(II)] \mbox{$s>1$}: The generalized connections~$\om{s}$ eat the
  fields $\om{s-1}$ as described by Eq.~(\ref{redef}). The
  connections~$\om{s-1}$ parameterize the coset space $\textit{Diff}_0(d,
  \RR)/GL(d,\RR)$, where $\textit{Diff}_0(d, \RR)$ is the homogeneous
  part of the diffeomorphism group. 
\end{itemize}
The absorption takes place in such a way that the fields
$\Ga{s}^\alpha{}_{\beta_1...\beta_s}$ ($s \geqslant 1$) and
$\Ga{1}_{(\alpha\beta)}$ ($\subset \Gamma_{G/H}$) turn into
rank-$s\textmd{+}2$ tensors, while $\Ga{1}_{[\alpha\beta]}$ ($=
\Gamma_{H}$) remains a true connection. Recall that the coset part
$\Gamma_{G/H}$ of the total connection transforms homogeneously under
the diffeomorphism group, while $\Gamma_H$ is a true connection.\footnote{
Note that, with $G$ being $\textit{Diff}(d,\mathbb{R})$ and $H$ being either
$GL(d,\mathbb{R})$ or $SO(1,d-1)$, the commutator of a generator of 
$G/H$ with a generator of $H$ is a linear combination of generators of $G/H$
(making $G/H$ a reductive coset), which ensures that $\Gamma_{G/H}$ and 
$\Gamma_{H}$ transform independently under $G$.} 

The coset fields $\om{s}$ play a fascinating double role at the
absorption process as can be seen by linearizing 
Eq.~(\ref{redef}):
\begin{align}
  \Ga{s}^\alpha{}_{\beta_1...\beta_{s}k} &=
  \partial_k \om{s\textmd{-}1}^\alpha{}_{\beta_1...\beta_s}  
  - (s+1) \, \om{s}^\alpha{}_{\beta_1...\beta_s k} \,. \label{redeflin}
\end{align}
Here $\om{s\textmd{-}1}$ behaves as a genuine Goldstone field which
gets absorbed by the field $\om{s}$.  The same field plays however a
different role on the next lower level. Considering
Eq.~(\ref{redeflin}) for $s-1$ instead of $s$, we see that
$\om{s\textmd{-}1}$ itself absorbs $\om{s\textmd{-}2}$.  In this
aspect it resembles more the characteristic behavior of a gauge boson.

The fact that in gravity Goldstone bosons can also take over the role
of absorber fields is related to the ``inverse Higgs effect'' 
\cite{Ivan}, see also \cite{Low} for a recent review. Goldstone's
theorem states that there is a {\em massless} mode for each broken
symmetry. However, since some of the Goldstone bosons can become
massive for spontaneously broken space-time groups \cite{Salam}, the
theorem gives only an upper bound on the number of massless Goldstone
modes.

The standard example is the spontaneous breaking of the conformal
group $SO(4,2)$ down to the Poincar\'e group $ISO(1,3)$ \cite{Salam,
Low}. {}From the dimension of the coset space one would expect five
massless Goldstone bosons, one corresponding to scale transformations
and four corresponding to special conformal transformations. However,
the special conformal parameter $\varphi_\mu$ becomes massive by the
absorption of the dilaton $\phi$ as can be seen from the total
nonlinear connection component along the dilations $\Gamma_{D}=
\varphi'_\mu dx^\mu=(2 \varphi_\mu - \partial_\mu \phi) dx^\mu$.

It is usually argued \cite{Ivan, Low} that one can set the part
$\Gamma_{G/H}$ of the total nonlinear connection $\Gamma$ to zero,
$\Gamma_{G/H}=0$. This gives relations among the coset fields, which
reduces the actual number of massless Goldstone fields.  For instance,
setting $\Gamma_D=0$ in the above example implies that $2\varphi_\mu$
can be replaced by $\partial_\mu \phi$. Note however that
$\Gamma_{G/H}=0$ should be interpreted as an effective equation, since
one ignores all massive Goldstone bosons ($\varphi'_\mu$ in the above
example). This constraint is justified only at energies much below the
mass of these Goldstone bosons.

In the realization considered in this paper, the relation
$\Gamma_{G/H}=0$ translates into $\Ga{s}=0$ $(s > 1)$ and
\mbox{$\Ga{1}_{(\alpha\beta)}=0$}. The constraint $\Ga{s}=0$ $(s > 1)$
relates all the generalized Goldstone connections by
\begin{align}
\om{s-1}_{[\alpha\beta_1]...\beta_{s-1}k}=\frac{2}{s!} \partial_{\beta_2}
...\partial_{\beta_{s-1}} \om{1}_{[\alpha\beta_1]k} \,.
\end{align}
We have shown in \cite{Kirsch} that by setting
\mbox{$\Ga{1}_{(\alpha\beta)}=0$}, the affine connection
$\Ga{1}_{[\alpha\beta]k}$ becomes metric-compatible, {\textit{i.e.}}\
equivalent to the Christoffel connection. In this way all higher spin
connections are given in terms of derivatives of the tetrad. We stress
again that this is only true at low energies, where all higher spin
fields are assumed to be decoupled.
%
\section{Higgs mechanism for gravity} \label{sec3}
%
The spontaneous breaking of the diffeomorphism group down to the
Lorentz group gives rise to an infinite tower of higher spin
connections as well as to the metric.  In this section we construct
actions for some parts of the corresponding Higgs mechanism by which the
higher spin connections get massive. Assuming their decoupling at low
energies, general relativity results as the appropriate effective
low-energy description of gravity.

\subsection{The breaking of dilations and shear transformations}
\label{secbreak}

One part of the symmetry breaking of the diffeomorphism group is the
spontaneous breaking of its linear subgroup
$GL(d,\RR)$~$\subset$~\textit{Diff}$(d,\RR)$ down to the Lorentz group
$SO(1,d-1)$. In the following we propose a Higgs mechanism for this
breaking which shows the occurrence of the metric as a Goldstone field
in an affine space-time. The model is largely based on that
of~\cite{Kirsch}. Previous Higgs models of the (special) linear group
have been constructed in \cite{Sija88, Kirsch2}.

We begin by assuming that all higher spin connections have already
been decoupled and we are left with a massless connection
$\Gamma^i{}_{jk}$ of an effective affine space-time. This connection
is considered as an independent dynamical variable and, in particular,
does not depend on the existence of the metric. In this space-time the
metric will be introduced as a Higgs field which breaks the linear
group $GL(d,\RR)$ in the tangent space.  Recall from Sec.~\ref{secIIb}
that the breaking of shear and dilation invariance leads to the metric
as a Goldstone field.

We construct the Higgs sector as follows.  In analogy to the complex
scalar $\Phi$ of $U(1)$ symmetry breaking, the breaking is induced by
a (real) scalar field $\phi$ and a symmetric tensor~$\varphi_{ij}$.
Under the Lorentz group the tensor $\varphi_{ij}$ decomposes into a
scalar $\sigma$ and a traceless symmetric tensor $\hat{h}_{ij}$ (${\bf
10 \rightarrow 1 + 9}$ in \mbox{$d=4$}),
\begin{align}
\varphi_{ij} = \hat h_{ij} + \frac 1 d \sigma \eta_{ij} \,,\qquad 
\eta^{ij}\hat{h}_{ij}\equiv 0\,.
\label{para1}
\end{align} 
The singlet $\sigma$ has been introduced in analogy to the Higgs field
in $U(1)$ symmetry breaking. 
The fields $h_{ij}=\hat{h}_{ij}+\frac{\phi}{d}\h_{ij}$ 
are the $d(d+1)/2$ Goldstone fields parameterizing the coset 
space $GL(d,\RR)/SO(1, d-1)$.
In fact, compared to the previous section, the coset fields $h_{ij}$
associated with the linear generators $T^{ij}$ have been rescaled 
and redefined so that they now possess mechanical dimension $m_P^{(d-2)/2}$.
The quantity $\k h_{ij}$ has no dimension, if $\k$ is the gravitational 
constant appearing in Einstein-Hilbert's action 
$S^{EH}[g_{ij}]=\frac{2}{\k^2}\int d^{d}x \,\sqrt{-g}\,R$. 
In terms of the Planck mass $m_{P}$, we thus have $\k=m_{P}^{(2-d)/2}$. 
Similarly, the quantity $\k\s$ is dimensionless, and so are 
$\k\varphi_{ij}$ and $\k\phi$.  
 
For later convenience, we also introduce
another parametrization for $\varphi_{ij}$, which is the analog of the
polar parametrization for the complex scalar field $\Phi$ in the
$U(1)$ Higgs mechanism:
\begin{align}
\varphi_{ij} &= \frac{\s}{d} e^{-\k\phi} g_{ij}\,,\quad
g_{ij} := e^{\k\phi}(e^{\k\bar{h}})_{ij}\,. 
\label{para2}  
\end{align}
In order for both parametrizations (\ref{para1}) and (\ref{para2}) to
define one and the same field $\varphi_{ij}$, it is easy to see that
the real, symmetric matrix $\bar{h}_{ij}$ must satisfy one
constraint\footnote{Being real and symmetric, $\bar{h}_{ij}$ can be
diagonalized by an orthogonal matrix $O$: $\bar{h}=ODO^{-1}$ where
$D=diag(\l_1,\ldots,\l_d)$.  Now, the identification $\hat h_{ij} +
\frac 1 d \sigma \eta_{ij}=\varphi_{ij}=$ $\frac{\s}{d}
(\exp[\bar{h}])_{ij}$ implies that $\hat{h}_{ij}=\frac{\s}{d}$
$[\bar{h}+\frac{1}{2!}\bar{h}^2+\frac{1}{3!}\bar{h}^3+ \ldots]_{ij}$.
Finally, because $\h^{ij}\hat{h}_{ij}\equiv 0$, one can easily see
that the above equation yields $\prod_{n=1}^d\l_n=\ln d$, that is, one
constraint on $\bar{h}$.}.  In other words, the matrix $\bar{h}$
possesses the same number of independent components as does $\hat{h}$,
{\textit{viz.}} $\frac{d(d+1)}{2} - 1$.  The fields $\phi$ and
$\bar{h}_{ij}$ may therefore as well parametrize the coset space
$GL(d,\RR)/SO(1, d-1)$.  The field $\phi$ parameterizes dilations
while $\bar{h}_{ij}$ parametrizes $SL(d,\RR)/SO(1, d-1)$.  We will use
the metric $g_{ij}$ and its inverse, denoted $g^{ij}$, to lower and
raise the indices. 

It is then convenient to define the {\em nonmetricity} tensor
$Q_{ijk}$ by
\begin{align} \label{defnonmet}
 Q_{ijk} \equiv -D_k g_{ij} = - \pr_k g_{ij} + 2 \Gamma^l{}_{k(i} g_{j)l} \,.
\end{align}
Note that this definition exactly reflects the symmetric part of the
absorption equation (\ref{coeff-2}) (identify $\Ga{1}_{(ij)k} \sim
Q_{ijk}$ and $\omega^i{}_{jk} \sim \Gamma^i{}_{jk})$. Solving this for
$\G^i{}_{jk}$,
\begin{align} \label{solution}
\Gamma^i{}_{jk}&={\Gamma}^{\{\}}{}^i{}_{jk} + N^i{}_{jk}\,, \\
N^i{}_{jk} &\equiv \frac{1}{2}(Q_{jk}{}^i +Q_k{}^i{}_j-Q^i{}_{jk}) \,,
\label{N}
\end{align}
we observe that a general symmetric connection can be expressed in
terms of the Christoffel connection
${\Gamma}^{\{\}}{}^i{}_{jk}(g_{ij})$ and nonmetricity.

We can now write down a $GL(d ,\RR)$ invariant action for the fields
$\G^i{}_{jk}(x)$, $\phi(x)$, $\varphi_{ij}(x)$ and the descendant
$g_{ij}(\phi(x), \hat h_{ij}(x))$. However, it turns out to be more
convenient to perform a change of variables by Eq.~(\ref{solution}),
$\G^i{}_{jk}(x) \rightarrow Q_{ijk}(x)$, and to construct instead an
action for $Q_{ijk}(x)$, $\phi(x)$, $\varphi_{ij}(x)$.  Nonmetricity
contains a totally symmetric part $\widetilde Q_{ijk} \equiv
Q_{(ijk)}$ which can be viewed as representing a massless spin-3
field, if no mass terms are introduced for $Q$.  We therefore have to
construct an action for a massive spin-2 field $\varphi_{ij}$ (and a
scalar $\phi$) in the background of a massless spin-3 field
$\widetilde Q_{ijk}$.

We are not aware of any action that would consistently couple a
massive spin-2 field to a massless spin-3 gauge field, but consistent
non-linear higher-spin field equations have been constructed
\cite{Vasiliev:2003ev}, that involve an infinite tower of higher-spin
gauge fields.

Our point of view in the present work is to postulate the existence of
an action in which all higher-spin gauge fields would consistently
interact, and focus only at particular sectors of this action.  Then,
those sub-sectors need not be separately consistent but must obey the
requirement that, in the free limit, they should reduce to a positive
sum of Singh-Hagen \cite{Singh:1974qz} and/or Fronsdal
\cite{Fronsdal:1978rb} actions.  Clearly, as we already mentioned,
the mechanism we are presenting here must be seen as a very small part
of a complete Higgs mechanism involving the infinite number of Goldstone
fields of the coset space $\textit{Diff}\,(d,\mathbb{R})/SO(1,d-1)$.

The action $S$ we propose is given by
\begin{align} \label{action}
S =  \int d^d x \sqrt{-g} \big[ \k^p{\cal L} (\widetilde Q_{ijk}) 
  + 3 {\cal L}(\phi, \varphi_{ij}) + {\cal L}_{\rm int} \big] \,,
\end{align}
with ($\widetilde Q_{i} \equiv \widetilde Q_{i}{}^k{}_k$, 
$p=\frac{2(d-4)}{2-d}$)
\begin{align} \label{LQ}
 {\cal L}(\widetilde Q_{ijk})= &-\frac 1 2(D_i \widetilde Q_{jkl})^2 
   + \frac{3}{2}(D^i \widetilde Q_{ijk})^2 + \frac{3}{2} 
   (D_i \widetilde Q_j)^2 \nonumber\\
   &+3 (D^ i D^j \widetilde Q_{ijk}) \widetilde Q^k + 
   \frac{3}4 (D^i \widetilde Q_i)^2 \,,\\
 {\cal L}(\phi, \varphi_{ij}) 
   &= \frac 1 2(D_i \varphi_{jk})^2 
   -\frac 1 2(D_i \varphi^k{}_k)^2
   - D_i\varphi_{jk} D^k\varphi^{ij} \nonumber\\
   &\quad+ D^i \varphi^k{}_k D^j\varphi_{ij} +\textstyle{\frac{3d-d^2-2}{2d^2}}
   (D_i \phi)^2 
\nonumber\\
   &\quad+ D^i \phi D^j \varphi_{ij} 
   - V(\phi,\varphi_{ij}) \,, \label{SSB} \\
 {\cal L}_{\rm int} =\,\,& \!\!- \varphi^{ij} \varphi_{ij} \widetilde Q^{klm}
   \widetilde Q_{klm} - \frac{3}{2d} (D^i \widetilde Q_{ijk}) \varphi^{jk}
   \varphi^{l}{}_l  
\end{align}
and the effective potential  
\begin{align}
 V(\phi, \varphi_{ij}) &= \frac{\lambda \k^p}{4} \left( \frac{\varphi^2}{\k^p} 
 - M^2 \right)^2 + \frac{m^2}{2} \phi^2 + \frac{\lambda'}{2\k^p}
 \varphi^2 \phi^2 \,, \nonumber \\
 \varphi^2 &\equiv (\varphi_{ij}g^{ij})^2 - \varphi^{ij} \varphi_{ij}\,.
 \nonumber
\end{align}

The kinetic terms in the Lagrangians ${\cal L} (\widetilde Q_{ijk})$
and ${\cal L}(\phi, \varphi_{ij})$ are obtained from the Fronsdal
Lagrangian for a massless spin-3 field $\widetilde Q_{ijk}$ and the
Fierz-Pauli Lagrangian for a massive \mbox{spin-2} field
$\varphi_{ij}$.  The Lagrangian ${\cal L}(\phi, \varphi_{ij})$
contains also a kinetic term for the scalar $\phi$ and a symmetry
breaking potential $V(\phi,\varphi_{ij})$. The kinetic terms in ${\cal
L}(\phi, \varphi_{ij})$ are invariant under the exchange of
$\varphi^k{}_k$ and $\phi$.  

By construction, the action is invariant under the linear group.  The
linear connection is minimally coupled to the fields $\phi$,
$\varphi_{ij}$ and $\widetilde Q_{ijk}$ via the covariant derivative
\begin{align}
 D_i \varphi_{jk} &= \partial _i \varphi_{jk} 
 - 2 \Gamma^l{}_{i(j}\, \varphi_{k)l}
= \nabla_i\varphi_{jk} - 2 N^l{}_{i(j}\, \varphi_{k)l} 
\end{align}
and similarly for $\widetilde Q_{ijk}$. Here $\nabla_i$ is the
covariant derivative constructed from the Christoffel connection
$\Gamma^{\{\}}{}^i{}_{jk}$ and $N^i{}_{jk}$ as in Eq.~(\ref{N}).
${\cal L}_{\rm int}$ contains some additional nonminimal interactions.

For brevity, we omitted kinetic terms for the field $\bar {Q}_{ijk}
\equiv Q_{ijk} -\widetilde Q_{ijk}$. $\bar {Q}_{ijk}$ enters the
Lagrangian via the covariant derivatives and is required for linear
invariance. Linear invariant actions for all components of the
nonmetricity can be found in \cite{MAG}. Note that $\bar {Q}_{ijk}$ is
nonpropagating in $d=4$ if massless.

The potential $V(\phi,\varphi_{ij})$ has a minimum at
\begin{align}
 v_\varphi^2 \equiv {\langle \varphi^2 \rangle}   =   \k^{p} M^2 -
 \frac{\lambda'}{\lambda}\phi^2 \,
\end{align}
and is invariant in the Goldstone direction that parameterizes
$SL(d,\RR)/SO(1,d-1)$. This can best be seen in the parameterization
(\ref{para2}) in which $V$, upon rescaling
$\sigma'=e^{-\k\phi}\sigma$, becomes identical to the potential of
hybrid inflation \cite{Linde}.  Scale invariance is softly broken at
energy scales of order of the parameter $M$ and below (we assume $m^2
\ll M^2$).

As in hybrid inflation, we assume that the dilaton field $\phi$ is
slow-rolling and large at the beginning of the breaking.  As long as
the dilaton $\phi$ is larger than the critical value $\phi^2_c=
\k^{p}\lambda M^2/ \lambda'$, the field $\varphi_{ij}$ is trapped at
$\varphi_{ij}=0$. The effective mass squared of $\varphi_{ij}$, 
\begin{align}
 m^2(\varphi_{ij})=\frac{\lambda'}{\k^p}\,\phi^2 - \lambda M^2 \,,
\end{align}
becomes negative as soon as the value of $\phi$ falls below $\phi_c$,
$\phi < \phi_c$, at which point the vacuum becomes meta-stable. Then
the field $\varphi_{ij}$ is not trapped at $\varphi_{ij}=0$ anymore
and rolls down the ``waterfall'' to its minimum $v_\varphi =
\pm\kappa^{p/2} M$ at $\phi_0=0$. In this way the breaking of scale
invariance triggers the spontaneous breaking of the special linear
group $SL(d,\RR) \subset GL(d,\RR)$ down to the Lorentz group.
%
\subsubsection*{Higgs phase and massive spin-3 fields}
%
We now study the action (\ref{action}) in the Higgs phase.  Below
Eq.~(\ref{defnonmet}), we identified the nonmetricity field $Q_{ijk}$
with the component $\Ga{1}_{(ij)k}$ of the total nonlinear connection
$\Gamma$.  Since $\Ga{1}_{(ij)k} \subset \Gamma_{G/H}$, we expect
$Q_{ijk}$ to acquire mass during the symmetry breaking. We thus have
to show that at the minimum of the potential the Lagrangian in
(\ref{action}) contains the Singh-Hagen Lagrangian for the massive
spin-3 field $\widetilde Q_{ijk}$.

Let us first verify that the Goldstone field $\hat h_{ij}$ becomes
massless at the minimum of the potential.  According to the general
Higgs procedure, we have to expand $\phi$ and $\varphi_{ij}$ around
the absolute minimum of the potential at $\phi_0=0$ and 
$v_\varphi \equiv \sqrt{\langle \varphi^2 \rangle} = \pm \k^{p/2}M$. 
We choose the parameterization
\begin{align}
  \varphi_{ij} &= \big({\textstyle\frac{v_\varphi}{\sqrt{d(d-1)}}}
  + \frac{1}{d} \s \big)\eta_{ij} +
  \hat h_{ij} ,\,\,\,\,  \phi =\phi_0 +
   \tilde \phi = \tilde \phi \,, \label{split} 
\end{align}
where the normalization is chosen such that
$\varphi^2=v_\varphi^2+...\,$.  Substituting this into the potential
$V(\phi, \varphi_{ij})$, we observe that the Goldstone field
$\hat{h}_{ij}$ is indeed massless, whereas the Higgs-like field
$\sigma$ obtains a positive mass,
\begin{align}
 m^2_\sigma \sim \lambda M^2 \frac{d-1}{d} \,.
\end{align}

It is much simpler to work in the unitary gauge 
\begin{align} \label{unitarygauge}
 \varphi_{ij} &= \big({\textstyle\frac{v_\varphi}{\sqrt{d(d-1)}}} +
  \frac{1}{d} \s \big) \eta_{ij} \,, \qquad \phi = 0 \,,
\end{align}
in which the Goldstone bosons $\hat h_{ij}$ and $\phi$ are gauged
away. This corresponds to the flat space limit $g_{ij} = \eta_{ij}$.
In this gauge the Lagrangian in (\ref{action}) reduces to
\begin{align} \label{result}
{\cal L}&=\k^p {\cal L}_{F}(\widetilde Q_{ijk}) + \frac{v_\varphi^2}{2}
\widetilde Q_{ijk}^2 - {\frac{3}{2} v_\varphi^2 (\widetilde Q^l{}_{lj})^2}
-  \frac{ 9}{4}\lambda M^2 \sigma^2 \nonumber\\
&\quad+ \frac{9}{16} (\pr_i  \sigma)^2 {-\frac
3 4 v_\varphi (\pr^i\widetilde Q^l{}_{li}) \sigma} + ... \,,
\end{align}
where dots denote additional mixed terms. The Fronsdal Lagrangian
${\cal L}_{F}(\widetilde Q_{ijk})$ in flat space follows directly from
${\cal L}(\widetilde Q_{ijk})$ in (\ref{LQ}), while the mass terms for
$\widetilde Q_{ijk}$ and its trace $\widetilde Q^l{}_{lj}$ descend
from the kinetic terms of the Goldstone bosons.  If we choose
$\lambda=1$, then Eq.~(\ref{result}) is nothing but the St\"uckelberg
Lagrangian for a massive spin-3 field $\widetilde Q_{ijk}$
\cite{Riccioni} which is equivalent to the spin-3 Singh-Hagen
Lagrangian \cite{Singh:1974qz}. The mass of $\widetilde Q_{ijk}$ is
given by
\begin{align}
m^2_Q &= \frac{1}{\k^{p}}\,v_\varphi^2 = M^2 \,. \label{massQ}
\end{align}
The vacuum expectation value $v_\varphi=\k^{p/2}M$ is a free parameter
in the model and has to be determined by experiment. If we assume that
our model is indeed related to hybrid inflation, we can make a rough
estimation of the mass $m_Q=M$. It has been found \cite{Watari} that
the parameter $M$ determined by the COBE normalization is roughly
$10^{15}-10^{16}$~GeV.

The development of the field $\sigma$ is quite exciting. Since we
introduced $\sigma$ as a Higgs-like field, we would have expected it
to be an independent massive scalar, just like the Higgs particle in
elementary particle physics. Instead, the field $\sigma$ turned out to
be the auxiliary scalar required in the Singh-Hagen Lagrangian for a
massive spin-3 field. We thus do not have an additional Higgs
particle.

In the general parameterization (\ref{split}), there are additional
terms involving the Goldstone metric. Let us assume that $m_Q^2$ is
very high such that $\widetilde Q_{ijk}$ and $\sigma$ decouple at low
energies. In this decoupling limit, the Lagrangian in (\ref{action})
effectively reduces to the linearized Einstein-Hilbert Lagrangian
${\cal L}_F(h_{ij})$ with $h_{ij}=\hat h_{ij} + \frac{1}{d} \phi
\eta_{ij}$.  Gravity is thus effectively described by general
relativity at the minimum of the potential.  This shows explicitly
that the condition $Q_{ijk} \equiv -D_k g_{ij} \sim \Ga{1}_{(ij)k}=0$
imposed by the ``inverse Higgs effect'' is an effective equation.
%
\subsection{Higgs mechanism for higher spin connections} \label{IIIB}
%
In the previous section we proposed a full Higgs mechanism for the
linear group including a symmetry breaking potential.  An essential
part of the mechanism was the absorption process~(I) of
Sec.~\ref{secIId}. A description of the breaking of the complete
diffeomorphism group appears to be quite complex. We therefore aim at
the more modest goal of modeling the absorption process~(II) of
Sec.~\ref{secIId} for $s=2$ without giving a symmetry breaking
potential. This is along the lines of \cite{Aragone, Aulakh, Bianchi}
which discuss a ``telescopic Higgs effect''. The latter effect is
briefly reviewed now.
%
\subsubsection{St\"uckelberg formalism and the ``telescopic Higgs effect''}
%
At the basis of the St\"uckelberg formalism lies the well-known fact
that $O(d-1)$, the little group for a massless particle in $d$$+$$1$
dimensions, is the same as the little group for a massive particle in
a $d\,$-dimensional space-time.  Consistent actions for massive
particles can indeed be obtained by dimensional reduction of massless
gauge-invariant actions.  The dimensionally reduced action is itself
invariant under a set of gauge invariances which display a ``telescopic
Higgs effect''.

For example, in the previous section we recovered the Singh-Hagen
Lagrangian for a massive spin-3 field as resulting from the
expression, in the unitary gauge, of a Lagrangian containing an
appropriate symmetry-breaking potential~$V$.  This Lagrangian can also
be obtained starting from Fronsdal's gauge-invariant Lagrangian 
${\cal L}_F(\phi_{MNP})$ for a massless spin-3 field $\phi_{MNP}$,
dimensionally reduced from $d+1$ down to $d$ dimensions. Upon dimensional reduction,
the field $\phi_{MNP}$ gives rise to the set of $d$-dimensional fields
$\{\phi_{ijk}, \phi_{ij},A_i,\psi \}$ entering a $d$-dimensional
St\"uckelberg Lagrangian ${\cal L}_S(\phi_{ijk}, \phi_{ij},A_i,\psi)$.
However, not all of these fields will survive.  The dimensionally
reduced action ${\cal L}_S(\phi_{ijk}, \phi_{ij},A_i,\psi)$ inherits
gauge invariances from ${\cal L}_F(\phi_{MNP})$ whose effect is to
eliminate all the fields but $\phi_{ijk}$ and $\phi$, the trace of
$\phi_{ij}$.
%
\subsubsection{Higher spin connections in the St\"uckelberg formulation}
\label{stu}
%
We now apply the St\"uckelberg formalism to model the absorption
process (II) of Sec.~\ref{secIId}. We begin by splitting
Eq.~(\ref{redeflin}) into irreducible pieces under the linear
group. The total symmetric and the ``hook'' parts are given by
\begin{align}
\Gahat{s}_{(\alpha\beta_1...\beta_{s}k)} &=
   \omhat{s\textmd{-}1}_{(\alpha\beta_1...\beta_s, k)}  
  - (s+1) \, \omhat{s}_{(\alpha\beta_1...\beta_s k)} \,, \nonumber\\
\Gahat{s}_{[\alpha(\beta_1]...\beta_{s}k)} &=
  \omhat{s\textmd{-}1}_{[\alpha(\beta_1]...\beta_s, k)}  
  - (s+1) \, \omhat{s}_{[\alpha\beta_1]...\beta_s k} \,,
\label{reducedabsorption}
\end{align}
where in the second line we first symmetrize in $\beta_1,...,\beta_s,
k$ and then antisymmetrize in $\alpha$ and $\beta_1$. A comma
denotes a partial derivative, e.g. $\Phi_{,k}:=\partial_k\Phi$.

We have restricted to (double) traceless fields, which is required for
the construction of Fronsdal's Lagrangians.  The hat on top of a field
indicates its tracelessness in the anholonomic indices, e.g.\
$\omhat{s}^{\alpha}{}_{\alpha\beta_2...\beta_s k}=
\omhat{s}_{\alpha}{}^\beta{}_{\beta\beta_3...\beta_s k}=0$. This
guarantees the double tracelessness of the field
$\omhat{s}_{\alpha\beta_1...\beta_s k}$. Note that the fully
anholonomic field
\begin{align}
 \omhat{s\textmd{-}1}_{\alpha\beta_1...\beta_s} = 
 \omhat{s\textmd{-}1}_{\alpha\beta_1...\beta_{s-1}k} \hat 
 e^k{}_{\beta_{s}}
\end{align} 
is traceless, rather than double traceless.  In the nonlinear
realization of $\textit{Diff}(d,\mathbb{R})$ with the Lorentz group
$SO(1,d-1)$ as stabilizing subgroup, it is indeed natural to decompose
the generators $F_{\a}^{~\b_1\ldots\b_2}$ of
$\textit{Diff}\,(d,\mathbb{R})$ --- as expressed in an anholonomic basis
--- with respect to irreducible representations of $SO(1,d-1)$.  In
other words, the generators $F_{\a}^{~\b_1\ldots\b_2}\equiv
{e}^i_{~\a}F_{i}^{~j_1\ldots j_s} {(e^{-1})}_{j_1}^{~\b_1}\ldots
{(e^{-1})}_{j_s}^{~\b_s}$ can be decomposed 
into their traceless and pure-trace parts by using the Minkowski metric
$\h_{\a\b}$.

It is convenient to define the fields
\begin{align}
 \phi_{\alpha\beta_1...\beta_s k} &\equiv
 \omhat{s}_{(\alpha\beta_1...\beta_s k)} \,, \nonumber \\
 \phi'_{\alpha\beta_1...\beta_s k} &\equiv
 \Gahat{s}_{(\alpha\beta_1...\beta_s k)} \nonumber \\
 T_{\alpha\beta_1|\beta_2...\beta_sk} &\equiv
 \omhat{s}_{[\alpha\beta_1]...\beta_s k} \,,\nonumber \\
 T'_{\alpha\beta_1|\beta_2...\beta_sk} &\equiv
 \Gahat{s}_{[\alpha(\beta_1]...\beta_s k)}
\end{align}
which, by assumption (more details in section \ref{secIVa}), 
satisfy the gauge transformation laws 
\begin{align}
\delta  \phi_{\alpha\beta_1...\beta_s}
&= s \partial_{(\alpha} \hat \lambda_{\beta_1...\beta_s)} \\
\delta T_{\alpha \beta_1 | \beta_2...\beta_s} &= \partial_{[\alpha} 
\hat k_{\beta_1]|\beta_2...\beta_s} \nonumber\\
&\,\,\,\,\,- \frac{3(s-1)}{s+1} \partial_{([\beta_2}
\hat k_{\alpha|\beta_1] \beta_3...\beta_s)} \,, \label{trafoT}
\end{align}
where the last term in (\ref{trafoT}) is first antisymmetrized in the
indices $\beta_2, \alpha, \beta_1$ and then completely symmetrized in 
the indices $\beta_2,\beta_3,...,\beta_s$ \cite{Boulanger:2003vs}. 
The gauge parameters
$\hat \lambda$ and $\hat k$ satisfy {$\hat \lambda^{\beta}{}_{\beta
\beta_3...\beta_s}=0$} and $\hat k_{\beta_1|}{}^{\beta}{}_{\beta
\beta_4...\beta_s}=0$.  
Using 
\begin{align}
\omega_{\a|\b_1\ldots\b_s} = 
\frac{2s}{s+1}T_{\a(\b_1|\b_2\ldots\b_s)}+\phi_{\a\b_1\ldots\b_s} \,,
\end{align}
Eq.~(\ref{reducedabsorption}) can then be rewritten as
\begin{align}
\phi'_{\alpha\beta_1...\beta_{s}k} &=
   \partial_{(k} \phi_{\alpha\beta_1...\beta_s)}  
  - (s+1) \, \phi_{\alpha\beta_1...\beta_s k} \,,\label{redabs1}\\
 T'_{\alpha\beta_1|\beta_2...\beta_{s}k} &=
 - (s+1) \, T_{\alpha\beta_1|\beta_2...\beta_s k}
 \nonumber\\
 &+\frac{2s}{(s+1)^2}\partial_{[\b_1}T_{\a](\b_2|\b_3\ldots\b_s k)} 
 \nonumber \\
 &+ \frac{s}{s+1}T_{\a\b_1|\b_2\ldots\b_s,k}  
   \nonumber\\
 &+\frac{1}{s+1}\partial_{[\b_1}\phi_{\a]\b_2\ldots\b_s k} \,.
\label{redabs2}
\end{align}

Eq.~(\ref{redabs1}) is the relevant gauge transformation involved in
the Higgs mechanism for totally symmetric spin-$s$+$2$ fields
$\phi_{\alpha\beta_1...\beta_s k}$ (Young tableau [s+2, 0]), while
Eq.~(\ref{redabs2}) describes the Higgs effect for spin-$s$+$1$ fields
$T_{\alpha\beta_1|\beta_2...\beta_sk}$ in the ``hook'' representation
[s+1, 1].

For simplicity, we restrict to $s=2$ in the following. 

Let us compare Eqs.~(\ref{redabs1}) and (\ref{redabs2}) with the fields
arising in the St\"uckelberg formalism. Decomposing the massive 
representations $[4,0]_m$ and $[3,1]_m$ into massless
representations, we obtain
\begin{align}
\overbrace{[4,0]}^{\phi'_{\alpha\beta_1\beta_2k}}\!\!\!{}_m 
&\rightarrow \overbrace{[4,0]}^{\phi_{\alpha\beta_1\beta_2k}} \!\oplus 
\,[3,0] \oplus [2,0] \oplus [1,0] \oplus [0,0]\\ 
\overbrace{[3,1]}^{T'_{\alpha\beta_1|\beta_2k}}\!\!\!\!{}_m &\rightarrow
\!\!\!
\overbrace{[3,1]}^{T_{\alpha\beta_1|\beta_2k}} \oplus
\overbrace{[2,1]}^{C_{\alpha\beta_1|\beta_2}} \oplus
\overbrace{[2,0]}^{\phi_{\alpha\beta}} \oplus 
\overbrace{[1,1]}^{B_{\alpha\beta}}
\oplus \overbrace{[1,0]}^{X_\alpha} \oplus
\overbrace{[3,0]}^{S_{\alpha\beta_1\beta_2}} \nonumber
\end{align} 
This shows that a massless representation $[4,0]$ has to absorb the
massless representations $[3,0],[2,0], [1,0], [0,0]$ to become
massive. These representations descend themselves from the massive
representation $[3,0]_m$ ($\phi_{\alpha\beta_1k}$). Moreover, the
massless representation $[3,1]$ must absorb the $[2,1], [2,0], [1,1],
[1,0]$ and $[3,0]$ to become massive.  The first four of these
representations come from the massive representation $[2,1]_m$
($T_{\alpha\beta|k}$). The remaining representation $[3,0]$ originates
from $\phi_{\alpha\beta_1k}$, cf.\ with the last line in
Eq.~(\ref{redabs2}).  

This agrees with the fact that, in a full Higgs mechanism including a
symmetry breaking potential, the Goldstone bosons
$\phi_{\alpha\beta_1k}$ and $T_{\alpha\beta|k}$ would be introduced
inside a (tachyonic) Higgs field, {\textit{i.e.}}\ as massive
representations. At the minimum of the potential the Goldstone bosons
condense and become massless. Note however that in the following we
restrict to give the St\"uckelberg description of the massive
representations which we consider as part of the full symmetry
breaking mechanism.
%
\subsubsection{St\"uckelberg formulation of massive [3,1] hook field}
%
The St\"uckelberg formalism for the representation \mbox{$[4,0]$} has
been discussed in detail in \cite{Bianchi}.\footnote{If fact, it is
discussed for all totally symmetric fields $\phi_{i_1 ...i_s}$ there.}
We therefore need to construct only a field equation for the
representation $[3,1]$ which describes the absorption (\ref{redabs2})
for $s=2$.

Upon dimensional reduction $(x^{i},y)\downarrow x^i$, the
$(d+1)$-dimensional massless gauge field $T_{MN|PQ}(x,y)$ gives rise
to the following $d$-dimensional gauge fields
\begin{align}
T_{MN|PQ}(x,y) &= \frac{1}{\sqrt{2}} T_{MN|PQ}(x)e^{imy} 
+ c.c.\,,
\nonumber \\
T_{ij|kl}(x,y) &= \frac{1}{\sqrt{2}} T_{ij|kl}(x)e^{imy} + c.c.\,,
\nonumber \\
T_{ij|k y}(x,y) &= \frac{i}{\sqrt{2}} C_{ij|k}(x)e^{imy} + c.c.\,,
\nonumber \\
T_{i y|jk}(x,y) &= \frac{i}{\sqrt{2}}[S_{ijk}(x)
+\frac{2}{3}\,C_{i(j|k)}(x)]e^{imy} + c.c.\,,
\nonumber \\
T_{ij |yy}(x,y) &= \frac{1}{\sqrt{2}}B_{ij}(x)e^{imy} + c.c.\,,
\nonumber \\
T_{i y|y j}(x,y) &= \frac{1}{\sqrt{2}}
[\phi_{ij}(x) +\frac{1}{2}B_{ij}(x)]e^{imy} + c.c.\,,
\nonumber \\
T_{i y|yy}(x,y) &= \frac{i}{\sqrt{2}}X_{i}(x)e^{imy} + c.c.\,.
\nonumber
\end{align}
The descendant fields $T_{ij|kl}, S_{ijk}, C_{ij|k}, B_{ij},
\phi_{ij}$ and $X_{i}$ are all real.  The field $T_{MN|PQ}(x,y)$ has
the following symmetries
\begin{align}
	T_{MN|PQ}&=-T_{NM|PQ}=-T_{NM|QP}\,,
  \quad T_{[MN|P]Q}\equiv 0\,,\nonumber 
\end{align}
while the descendant $d$-dimensional fields have the symmetries 
\begin{align}
C_{ij | k} &= -C_{ji|k}\,,\quad C_{[ij|k]}\equiv 0\,,
\nonumber \\
S_{ijk}&= S_{(ijk)}\,,~B_{ij}=-B_{ji}\,,~\phi_{ij}=\phi_{ji}\,.
\nonumber
\end{align}
The gauge transformations of the field $T_{MN|PQ}(x,y)$ are
\begin{align}
\delta T_{MN|PQ} &= \pa_{[M}\hat{K}_{N]|PQ}
	\nonumber \\
	&~-\frac{3}{4}\left(\pa_{[P}\hat{K}_{M|N]Q}
          +\pa_{[Q}\hat{K}_{M|N]P}\right)  \,,
\nonumber
\end{align}
where the gauge parameter $\hat{K}_{N]|PQ}(x,y)$ possesses the
following symmetries
\begin{align}
	\hat{K}_{M|NP} &= \hat{K}_{M|PN}\,,\nonumber \\  
	0 &= \hat{K}^{~~~~\;N}_{M|N} \equiv \hat{K}_{M|\a}^{~~~~\a}+\hat{K}_{M|yy} \,. 
\nonumber
\end{align} 
The $(d+1)$-dimensional gauge parameter $\hat{K}_{M|NP}(x,y)$
generates the following $d$-dimensional gauge parameters upon
dimensional reduction
\begin{align}
	\hat{K}_{M|NP}(x,y) &= \frac{1}{\sqrt{2}}\hat{K}_{M|NP}(x)e^{imy}+ c.c.\,,
\nonumber \\
	\hat{K}_{i|jk} &= \frac{1}{\sqrt{2}}[
	\hat{k}_{i|jk}+\frac{\h_{jk}}{d}a_{i}]e^{imy}+c.c.\,,
\nonumber \\
  \hat{K}_{i|j y} &= \frac{i}{\sqrt{2}}\;{t}_{ij} e^{imy}+c.c.
\nonumber \\
  \hat{K}_{y|ij} &= \frac{i}{\sqrt{2}}
  [{\hat l}_{ij}+\frac{\h_{ij}}{d} l ]e^{imy}+c.c.\,,
\nonumber \\
  \hat{K}_{y|y i} &= \frac{1}{\sqrt{2}} \,\e_{i}\,e^{imy}+c.c.\,,
\nonumber \\
   \hat{K}_{y|yy}&= \frac{-i}{\sqrt{2}}\, l \,e^{imy} + c.c.\,,
\nonumber
\end{align} 
where the descendant gauge parameters $\hat{k}_{i|jk}$, $a_{i}$, 
${t}_{ij}$, ${\hat{l}}_{ij}$, $l$ and $\e_{i}$ are all real and 
\begin{align}
a_{i} &= \hat{K}_{i|k}^{~~~\,k}(x)\,,
\quad l = \hat{K}_{y|k}^{~~~\,k}(x)\,.\nonumber
\end{align}
They furthermore obey 
\begin{align}
\hat{k}_{i|jk}&=\hat{k}_{i|kj}\,,\quad \h^{jk}\hat{k}_{i|jk}=0
\nonumber \\
\hat{l}_{ij}&=\hat{l}_{ji}\,,\quad \h^{ij}\hat{l}_{ij}=0\,.
\nonumber
\end{align}
\subsubsection{Gauge transformations and field redefinitions}
%
The $d$-dimensional gauge transformations of the descendant fields read 
\begin{align}
	\delta T_{ij|kl} &= \pa_{[i}\hat{k}_{j]|kl}-
	\frac{3}{4}\left(\pa_{[k}\hat{k}_{i|j]l}+\pa_{[l}\hat{k}_{i|j]k}
	\right)
	\nonumber \\
	&+\frac{1}{2d}\h_{kl}\pa_{[i}a_{j]}-\frac{1}{4d}\left(
	\h_{jl}\pa_{[k}a_{i]}+\h_{jk}\pa_{[l}a_{i]}\right.
	\nonumber \\
	& \left. +\h_{il}\pa_{[j}a_{k]}+\h_{ik}\pa_{[j} a_{l]}\right)\,,
\nonumber \\
	\delta C_{ij|k} &= \frac{3}{4} (\pa_{[i}t_{j]k}-\pa_{[k}t_{ij]})
	-\frac{m}{4}\Big[\hat{k}_{[i|j]k}+\frac{1}{d}\h_{k[j}a_{i]}\Big]
	\nonumber \\
	&+\frac{1}{4}\pa_{[i}\hat{l}_{j]k}+\frac{1}{4d}\h_{k[j}\pa_{i]}l\,,
\nonumber \\
	\delta S_{ijk} &= \frac{1}{2}\pa_{(i}\hat{l}_{jk)}+
	\frac{1}{2d}\h_{(ij}\pa_{k)} l 
\nonumber\\
	&\quad\quad-\frac{m}{2}\Big(\hat{k}_{(i |jk)} + \frac{1}{d}\h_{(ij}a_{k)}\Big) \,,
\nonumber 
\end{align}
\begin{align}
\delta B_{ij} &= \frac{1}{2}\left( \pa_{[i} \e_{j]}-\pa_{[i} a_{j]} 
+m t_{[i j]}\right)\,,
\nonumber \\
\delta\phi_{ij} &= \frac{1}{2}\left( \pa_{(i} \e_{j)}+m t_{(i j)}  \right)\,,
\nonumber \\
\delta X_{i} &= \frac{1}{2}\left(m a_{i}  - \pa_{i} l \right)\,. 
\nonumber
\end{align}
The field $X_{i}$ drops out of the action 
by doing the following field redefinitions
\begin{align}
	\phi_{ij}&\longrightarrow \phi'_{ij} = \phi_{ij}\,,
	\label{phired1}\\
	B_{ij} &\longrightarrow B'_{ij} = B_{ij}+\frac{1}{m}\pa_{[i}X_{j]}\,,
	\label{Bred1}\\
	S_{ijk} &\longrightarrow S'_{ijk} = S_{ijk}+\frac{1}{D}\h_{(ij}X_{k)}\,,
	\label{Sred1}\\
	C_{ij | k} &\longrightarrow C'_{ij | k} 
	= C_{ij | k}+\frac{1}{2D}\h_{k[j}X_{i]}\,,
	\label{Cred1}\\
	T_{ij | k\s} &\longrightarrow
	T'_{ij | k\s} = T_{ij | k\s}
	\nonumber \\
	&+\frac{3}{2mD}\left[ 
	\h_{(jk}\pa_{\s)}X_{i}+\h_{i(j}\pa_{k}X_{\s)} \right.
	\nonumber \\
	&\left. - \h_{(ik}\pa_{\s)}X_{j}-\h_{j(i}\pa_{k}X_{\s)}\right]\,.
	\label{Tred1}
\end{align}
The redefined fields transform as
\begin{align}
\delta T'_{ij|kl} &= \pa_{[i}\hat{k}_{j]|kl}-
	\frac{3}{4}\left(\pa_{[k}\hat{k}_{i|j]l}+\pa_{[l}\hat{k}_{i|j]k}
	\right)\,,
 \label{deltaT1} \\
	\delta C'_{ij|k} &= \frac{3}{4}\left( \pa_{[i}t_{j]k}-\pa_{[k}t_{ij]}\right)
	-\frac{m}{4}\,\hat{k}_{[i|j]k}+\frac{1}{4}\pa_{[i}\hat{l}_{j]k}\,,
	\label{deltaC1}\\
\delta S'_{ijk} &= \frac{1}{2}\pa_{(i}\hat{l}_{jk)} - 
	\frac{m}{2}\,\hat{k}_{(i |jk)}  \,,
 \label{deltaS1}\\
\delta B'_{ij} &= \frac{1}{2} \left( 
\pa_{[i} \e_{j]} + m\, t_{[i | j]} \right) \,,
\nonumber \\
\delta \phi'_{ij}&= \frac{1}{2}\left( \pa_{(i} \e_{j)}
+ m\, t_{(i | j)}\right)\,.
\nonumber
\end{align}
Performing the field redefinitions (\ref{phired1})--(\ref{Tred1}) is
equivalent to going in the gauge
\begin{eqnarray}
	a_{i}=\frac{1}{m}\,\pa_{i}l\,,
	\label{gauge1}
\end{eqnarray} 
whose effect is to eliminate $X_{i}$ from the action.  Of course we
must have $m\neq 0$.  There is no redefined field varying with respect
to the gauge parameters $a_{i}$ and $l\,$.

\noindent The next gauge fixing condition that we choose is ($m\neq 0$)
\begin{eqnarray}
	t_{i j} = -\frac{1}{m}\,\pa_{i} \e_{j}
	\label{gauge2}
\end{eqnarray}
which is equivalent to gauging $B_{ij}$ and $\phi_{ij}$ away.  Note
that (\ref{deltaT1}), (\ref{deltaC1}) and (\ref{deltaS1}) are
unaffected by this gauge fixing condition.  In terms of field
redefinition, the gauge (\ref{gauge2}) translates as
\begin{align}
T'_{ij | kl} \longrightarrow T''_{ij | kl}&=T'_{ij | kl}
\nonumber \\
	C'_{ij | k}\longrightarrow C''_{ij | k}&=C'_{ij | k}
	-\frac{3}{2m}\,\pa_{[i}\phi'_{j]k}
	\nonumber \\
	&-\frac{1}{2m}\left(
	\pa_{k}B'_{ji}+\pa_{[i}B'_{j]k}\right)
\nonumber \\
S'_{ijk} \longrightarrow S''_{ijk}&=S'_{ijk}\,.
\nonumber
\end{align}
As a result, the fields $B'$ and $\phi'$ disappear from the action and 
we have
\begin{align}
\delta T''_{ij|kl} &= \pa_{[i}\hat{k}_{j]|kl}-
	\frac{3}{4}\left(\pa_{[k}\hat{k}_{i|j]l}+\pa_{[l}\hat{k}_{i|j]k}
	\right)\,,
\nonumber \\
	\delta C''_{ij|k} &= \frac{1}{4}\left(
	\pa_{[i}\hat{\l}_{j]k}-m\, \hat{k}_{[i|j]k}\right)\,,
\nonumber \\
\delta S''_{ijk} &= \frac{1}{2}\left( \pa_{(i}\hat{\l}_{jk)} - 
	{m}\,\hat{k}_{(i |jk)}  \right)\,.
\nonumber
\end{align}
Obviously, the next gauge condition we impose is ($m\neq 0$)
\begin{eqnarray}
	\hat{k}_{i | jk} = \frac{1}{m}\,\pa_{i}\hat{l}_{jk}
	\label{gauge3}
\end{eqnarray}
which enables us to eliminate the $(jk)$-\emph{traceless} part of 
$P_{i | jk}:=S''_{ijk}+\frac{8}{3}\,C''_{i(j|k)}$ since it can be
seen that 
\begin{eqnarray}
\delta P_{i| jk} = \frac{1}{2}\left(\pa_{i}\hat{l}_{jk}-m\hat{k}_{i|jk}\right)\,. 	
\nonumber
\end{eqnarray}
In other words, we can gauge away $S''_{ijk}$ and $C''_{ij| k}$ (the
two independent components of $P_{i|jk}$) \emph{except for the trace}
$\h^{jk}(S''_{ijk} + \frac{8}{3}\,C''_{ij|k})$
$=:S''_{i}+\frac{8}{3}\,C''_{i}$ which will remain in the action,
playing the r\^ole of an auxiliary vector field $V_{i}$ that we need
for the action of a massive $[3,1]$ field.  At the level of the
action, the gauge (\ref{gauge3}) translates as the field redefinition
\begin{align}
T''_{ij|kl} &\longrightarrow H_{ij|kl}=T''_{ij|kl}
\nonumber \\
&+ \frac{2}{m}\left[
\pa_{[i}{P}_{j]|kl}-
	\frac{3}{4}\left(\pa_{[k}{P}_{i|j]l}+\pa_{[l}{P}_{i|j]k}
	\right)
\right]\,.
\nonumber
\end{align}
This equation expresses the absorption (\ref{redabs2}). Then, at the
end of all these field redefinitions which are the translation of the
gauge-fixing conditions (\ref{gauge1}), (\ref{gauge2}) and
(\ref{gauge3}), all the fields but $H_{ij|kl}$ and $V_{i}:=
S''_{i}+\frac{8}{3}\,C''_{i}$ remain in the action.  The field
$H_{ij|kl}$ does not transform anymore, it has become a massive field.
The field $V_{i}:= S''_{i}+\frac{8}{3}\,C''_{i}$ does not transform
neither, it is an auxiliary field, as we show explicitly in the
following.
%
\subsubsection{Field equations}
%
The field equations for a massless $[3,1]$ irreducible hook field
$T_{MN|RS}$ in dimension $(d+1)$ are
\cite{Bekaert:2003az,Boulanger:2003vs} (see also \cite{Chung:1987mv}
in different symmetry conventions)
\begin{eqnarray}
F_{MN|AB}=0\,,	
\end{eqnarray}
where $F_{MN|AB}$ is the kinetic tensor
\begin{align}
	F_{MN|AB}&= \pa_R\pa^R T_{MN|AB} + 2\pa^R\pa_{[M}T_{N]R|AB} 
	\nonumber \\
	&-2 \pa^R T_{MN|R(A,B)} 
	+4 \pa_{[M}T^{R}_{~\;N]|R(A,B)} 
	\nonumber \\
	&+ \pa_A\pa_B T^{~~~~\;R}_{MN|~R}\,.  
	\nonumber
\end{align}
As before, a coma $\Phi_{B,A}$ denotes a partial derivative $\pa_A\Phi_B$. 
\noindent Partially decomposing the field equation according to $x^M=(x^i,y)$ gives
\begin{align}
	0&=\Box T_{MN|AB}+\pa_y\pa_y T_{MN|AB} - \pa_M \pa^{k}T_{k N|AB} 
	\nonumber \\
	&- \pa_M\pa_y T_{yN|AB} + \pa_N\pa^{k}T_{k M|AB} + \pa_N\pa_y T_{yM|AB}
	\nonumber \\
	&- \pa_A\pa^{k}T_{MN|k B}-\pa_A\pa_y T_{MN|yB}-\pa_B\pa^{k}T_{MN|k A}
	\nonumber \\
	&-\pa_B\pa_y T_{MN|yA}+\pa_A\pa_M T^{k}_{~N|k B} + \pa_A\pa_M T_{yN|yB}
	\nonumber \\
	&-\pa_A\pa_N T^{k}_{~M|k B}- \pa_A\pa_N T_{yM|yB} + \pa_B\pa_M T^{k}_{~N|k A} 
	\nonumber \\
	&+ \pa_B\pa_M T_{yN|yA}-\pa_B\pa_N T^{k}_{~M|k A} - \pa_B\pa_N T_{yM|yA}
	\nonumber \\
	&+\pa_A\pa_B T^{~~~~k}_{MN|~k} + \pa_A\pa_B T_{MN|yy}\,.
	\nonumber
\end{align}
We can now decompose the above expression where the indices $MN|AB$ take the values 
$ij|ab$, $i y|ab$, $ij|yy$, $i y| yb$ and $i y|yy$, respectively. We find
\begin{align}
	0&=F_{ij |ab}-m^2 T_{ij | ab}-2m\pa_{[i}S_{j]ab}-\frac{4}{3}m\,\pa_{[i}C_{j](a|b)}
	\nonumber \\
	&+\pa_a\pa_b B_{ij} + 2m C_{ij|(a,b)} - 4\pa_{[i}\phi_{j](a,b)}
	-2\pa_{[i}B_{j](a,b)}\,,\nonumber \\
	0 &= \Box(S_{i ab}+\frac{2}{3}C_{i(a|b)}) -\pa_{i}\pa^{k}
	(S_{k ab}+\frac{2}{3}C_{k(a|b)}) 
	\nonumber \\
	&- \pa^{k}\pa_a (S_{bik}+\frac{2}{3}C_{i(k |b)})
	- \pa^{k}\pa_b (S_{aik}+\frac{2}{3}C_{i(k |a)})
	\nonumber \\
	&+\;\pa_a\pa_b (S^{k}_{~ik}+\frac{2}{3}C^{~\;k}_{i ~\, | k})
	+m\pa^{k}T_{k i |ab}
	\nonumber \\
	&-2m T^{k}_{~\;i | k(a,b)} +\pa_a\pa_bX_{i}\,,
	\nonumber \\
	0&= \Box (\phi_{i b}+\frac{1}{2}B_{i b})-\pa_b\pa^{k}(\phi_{i b}
	+\frac{1}{2}B_{i b})
	\nonumber \\
	&+\pa_b\pa_{i}\phi^{k}_{~\,k}+m\pa^{k}S_{k i b} 
	-2m\pa_{(i}S_{b)k}^{~~\;\;k}-m\pa_b X_{i} 
	\nonumber \\
	&-\pa_{i}\pa^{k}(\phi_{k b}
	+\frac{1}{2}\,B_{k b})-\frac{4}{3}m\,\pa^{k}C_{k i | b} 
	+\frac{1}{3}m\,\pa^{k}C_{i b| k}
	\nonumber \\
	&-\frac{1}{3}m\,\pa_{i}C^{k}_{~\,b| k}
	-\frac{5}{3}m\,\pa_{b}C^{~\,k}_{i ~| k}
	+m^2 T^{k}_{~\;i | k b}\,,
	\nonumber  \\
	0&= \Box X_{i}-\pa_{i}\pa^{k}X_{k}- m^2 X_{i}+2m\pa^{k}B_{k i}
	\nonumber \\
	&-2m\pa^{k}\phi_{ik}+2m\pa_{i}\phi^{k}_{~\,k}
	-m^2\left(S_{ik}^{~~k}+\frac{8}{3}C^{~\,k}_{i~\, |k}\right)\,,
	\nonumber
\end{align}
where $F_{ij |ab}$ is the kinetic tensor for the field $T_{ij |ab}$. 

\noindent We now perform all the field redefinitions given in the
previous section.  The above field equations read
\begin{align}
 0&=F_{ij |ab}(H)-m^2 H_{ij |ab}
 \nonumber \\
 &-\frac{2}{m}\,\pa_a\pa_b\pa_{[i}\left( 
 {S''}_{j]k }^{~~~k }+\frac{8}{3}\,{C''}_{j]k  |}^{~~~\;k }\right)\,,
 \nonumber \\
 0&=\pa_a\pa_b\left( 
 {S''}_{ik }^{~~~k }+\frac{8}{3}\,{C''}_{ik  |}^{~~~\;k }\right)
 \nonumber \\
 &+m\left(\pa^{k }H_{k i  |ab}-\pa_a H_{k i  | b}^{~~~~\,k }-\pa_b H_{k i  | a}^{~~~~\,k }
 \right)\,,
 \nonumber \\
 0&= m^2 H_{k i  | b}^{~~~~\,k } - m \,\pa_b \left( 
 {S''}_{ik }^{~~~k }+\frac{8}{3}\,{C''}_{ik  |}^{~~~\;k }\right) \,,
 \nonumber \\
	0&=m^2 \Big[{S''}_{ik }^{~~~k }+\frac{8}{3}\,{C''}_{ik  |}^{~~~\;k }\Big]\,.
	\nonumber
\end{align}
All together, these field equations imply
\begin{eqnarray}
	(\Box - m^2) H_{ij | kl} &=&0\,,
	\nonumber \\
	\pa^{i} H_{ij | kl} &=& 0 \,,
	\nonumber \\
	\h^{jk} H_{ij | kl} &=& 0 \,,
	\nonumber
\end{eqnarray}
which are the field equations for a massive $D$-dimensional $[3,1]$ hook field. 
We thus derived the correct field redefinitions which express the absorption 
phenomenon by which a massless $[3,1]$ hook field becomes massive. 
%
\section{Geometrical interpretation of higher spin connections} 
\label{sec4}
%
In Sec.~II we found an infinite tower of generalized connections
$\om{s}$ parameterizing the coset space $\textit{Diff}_0(d,
\RR)/GL(d,\RR)$ associated with nonlinear coordinate
transformations. In Sec.~\ref{secIVa} we will relate them with higher
spin connections known from the frame formalism of higher spin
fields \cite{Vasiliev:1980as}. 
In particular, we derive some gauge invariance principle for
the generalized connections which leads to a geometrical
interpretation of higher spin connections.  In Secs.~\ref{secIIIb} and
\ref{secIIIc} we then study the geometrical structure of a space-time
equipped with higher spin connections.
%
\subsection{Gauge transformations of higher spin connections} 
\label{secIVa}
%
In the frame formalism for higher spin gauge fields in Minkowski space 
\cite{Vasiliev:1980as}, 
Lorentz-like connections $\omega_{k}{}^{\alpha|\beta_1...\beta_{S-1}}$ 
are given in terms of frame-like fields ${e}_k{}^{\beta_1...\beta_{S-1}}$. 
These fields are symmetric in the indices 
$\beta_1,...,\beta_{S-1}$ and satisfy the relations 
\begin{align}
\omega_{k|}{}^{\beta}{}_{|\beta\beta_2...\beta_{S-1}}&=0\,,\quad
\omega_{k|\alpha|}{}^{\beta}{}_{\beta\beta_3...\beta_{S-1}}=0 \,,
\nonumber \\
\omega_{k|(\alpha|\beta_1...\beta_{S-1})}&=0 \,, \quad
{e}_k{}^{\beta}{}_{\beta}{}^{\beta_3...\beta_{S-1}}=0\,.
\nonumber
\end{align} 
The higher spin connections and tetrads are invariant under
the gauge transformations 
\begin{align} \label{genLorentz}
\delta \omega_{k|\alpha|\beta_1...\beta_{S-1}} &= \partial_{k} 
a_{\alpha|\beta_1...\beta_{S-1}} + \Sigma_{k|\alpha | \beta_1...\beta_{S-1}} \,,\\
\delta {e}_{k|\beta_1...\beta_{S-1}} &= \partial_{k} 
\lambda_{\beta_1...\beta_{S-1}} + a_{k|\beta_1...\beta_{S-1}} \,,
\nonumber
\end{align}
where the gauge parameters $a$, $\Sigma$ and $\lambda$ 
are traceless, completely symmetric in the indices 
$(\beta_1...\beta_{S-1})$ and possess the following 
supplementary symmetry properties: 
\begin{align}
&a_{(\alpha|\beta_1...\beta_{S-1})} = 0 = 
\Sigma_{k|(\alpha | \beta_1...\beta_{S-1})}\,,
\nonumber \\
&\Sigma_{k|\alpha | \beta_1...\beta_{S-1}}=\Sigma_{\a| k | \beta_1...\beta_{S-1}}\,.
\nonumber
\end{align}
Of course, similar gauge transformation formulas are also present in the
metric-like formulation of higher spin gauge fields \cite{Fronsdal:1978rb} 
(see also \cite{deWit})
and are crucial for the construction of consistent higher spin theories.

Though nonlinear realizations are different from gauging, the group
action on the coset fields is very similar to a gauge transformation
\cite{Salam}. We may exploit this similarity to derive a gauge
transformation for the generalized connections $\om{s}$
which is basically given by Eq.~(\ref{genLorentz}).

We begin by rewriting the defining equations of the nonlinear
connection one-forms $\Ga{s}$. 
For simplicity, we consider once again the linearized version 
of Eq.~(\ref{redef}):
\begin{align}
\delta \om{s}^\alpha{}_{\beta_1...\beta_sk}
&= \partial_k \om{s-1}^\alpha{}_{\beta_1...\beta_s}\,, \quad(s>1)
\label{gaugetrafo2}
\end{align}
where the variation $\delta \om{s}{}^\alpha{}_{\beta_1...\beta_sk}$
has been defined by
\begin{align}
\delta \om{s}{}^\alpha{}_{\beta_1...\beta_sk} &\equiv
\Ga{s}{}^\alpha{}_{\beta_1...\beta_sk} - (-s-1) 
\om{s}^\alpha{}_{\beta_1...\beta_sk}  \nonumber
\end{align}
and where one takes the traceless projection of this equation in the 
anholonomic indices, as we did in Sec.~\ref{stu}. 
For the interpretation of Eq.~(\ref{gaugetrafo2}) as the gauge
transformation of the coset field $\om{s}$, we have to consider the
field $\om{s-1}^\alpha{}_{\beta_1...\beta_s}$ as the gauge parameter
of $\om{s}$. Indeed, if we define $a_{[\alpha|\beta_1]\b_2...\beta_s} :=
\om{s-1}_{[\alpha\beta_1]\b_2...\beta_s}$ and
$\omega_{k|[\alpha|\beta_1]\b_2...\beta_s} := 
\om{s}_{[\alpha\beta_1]\b_2...\beta_s k}$, then 
Eq.~(\ref{gaugetrafo2}) antisymmetrized in $(\a\b_1)$ 
is equivalent to the transformation (\ref{genLorentz}) in 
the manifestly antisymmetric conventions.\footnote{Since
$\om{s}{}^\alpha{}_{\beta_1...\beta_sk}$ is already completely
symmetric in the indices $({\beta_1...\beta_sk})$, there is no further
parameter $\Sigma_{k|\alpha|\beta_1...\beta_s}$ on the right-hand-side
of Eq.~(\ref{gaugetrafo2}).} 

It is crucial to observe here that a~certain coset field $\om{s}$
plays simultaneously the role of a gauge field as well as that of a
gauge parameter: On the one hand, the field
$\om{s-1}^\alpha{}_{\beta_1...\beta_s}$ acts as the gauge parameter of
the connection $\om{s}^\alpha{}_{\beta_1...\beta_sk}$. On the other
hand, on the next higher level,
$\om{s}^\alpha{}_{\beta_1...\beta_s\beta_{s+1}}$ has to be interpreted
itself as the gauge parameter of
$\om{s\textmd{+}1}^\alpha{}_{\beta_1...\beta_{s+1}k}$.  We have
already encountered this double role in Sec.~\ref{secIId}, where we
interpreted Eq.~(\ref{redef}) as an absorption equation.

Why do we expect Eq.~(\ref{redef}) to reproduce the gauge
transformations of higher spin connections? In Sec.~\ref{secIId} we
interpreted Eq.~(\ref{redef}) as an absorption equation for Goldstone
bosons. In the standard Higgs mechanism of elementary particle physics
the absorption of a Goldstone boson by a gauge field is identical to a
gauge transformation in which the gauge parameter is identified with
the Goldstone boson.  It is thus natural to regard Eq.~(\ref{redef})
as a kind of gravitational gauge transformation.

Continuing the analogy to gauging even further, we may ask which
global symmetry is made local by the generalized connections. Note
that the fields $\om{s}^{\alpha}{}_{\b_1...\b_{s}k}$ are the
components of the connection one-forms
$\Gamma^\alpha{}_{\b_1...\b_{s}}$ associated with the generators
$F^{(s-1)}_\alpha{}^{\b_1...\b_{s}}$ ($s\geqslant 1$). In this sense,
the global symmetries generated by $F^{(s-1)}_\a{}^{\b_1...\b_s}$ are
``gauged'' by $\om{s}^\alpha{}_{\beta_1...\beta_{s}k}$. For $s=1$ this
implies that the ordinary connection $\omega^{\alpha}{}_{\beta k}$ is
the gauge potential of the linear group.
%
\subsection{The strong equivalence principle} \label{secIIIb}
%
Gravity in a spacetime equipped with generalized connections obeys
the strong equivalence principle (SEP). The SEP states that
gravitational interactions can be gauged away by an appropriate
coordinate transformation. To see this, we prove that at each point
$P$ there exists a coordinate system such that
\begin{align}
\om{s}^i{}_{j_1...j_s k} \vert_P = 0 \,
\nonumber
\end{align}
for all $s \geqslant 1$.  

Let us choose $P$ as the point of origin $x^i=0$ (choose gauge
$\xi^i=x^i$) and perform the coordinate transformations
\begin{align} 
\label{xtrafo}
x^i \rightarrow x'{}^i = x^i + \frac{1}{(s+1)!}
\varepsilon^i{}_{j_1...j_sk} x^{j_1}\cdots x^{j_s}x^{k} \,.
\end{align}
Substituting this into the transformation law (\ref{finitetrafo}), we
obtain\footnote{We perform the coordinate transformation
(\ref{xtrafo}) first for $\om{1}$, then for $\om{2}$, etc. In this way
the term ${\cal O}(\om{s\textmd{-}1})$ in (\ref{finitetrafo}) is
absent, since we have already set all lower spin connections to zero.}
\begin{align} 
\om{s}'{}^i{}_{j_1...j_sk}\vert_P = \om{s}^i{}_{j_1...j_sk}\vert_P - 
\varepsilon^i{}_{j_1...j_s k} \,.
\nonumber
\end{align}
If we choose the parameters $\varepsilon^i{}_{j_1...j_sk}=
\om{s}^i{}_{j_1...j_sk}\vert_P$, we get $\om{s}'{}^i{}_{j_1...j_s
k}\vert_P = 0$ and, from this, $\Ga{s}{}^i{}_{j_1...j_s k}\vert_P = 0$
for all $s \geqslant 1$. It is thus possible to find a coordinate
system at a point~$P$ in which there is no gravitational force on a
point particle, {\textit{i.e.}}\ ${\ddot x^i}\vert_P~=~0$ (SEP).  All
higher spin connections have been gauged away.
%
\subsection{Velocity-dependent affine connection} \label{secIIIc}
%
There exists an interesting alternative view of a spacetime endowed
with higher spin connections. This view is based on a geometrical
object called {\em $N$-connection} ($N$ for nonlinear).  The concept
of an $N$-connection $N^i{}_{j} (x,\dot x)$ was first introduced by
\'E.~Cartan \cite{Cartan} in his work on Finsler spaces, see
\cite{Vacaru} for a modern review.  The $N$-connection is related to a
velocity-dependent affine connection $\gamma^i{}_{jk}(x,\dot x)$ by
\begin{align} \label{velconn}
N^i{}_{j} (x,\dot x) = \frac{1}{2} \frac{\partial}{\partial \dot x^j}
\left( \gamma^i{}_{nk}(x,\dot x)\, \dot x^n \dot x^k \right) \,.
\end{align}

The affine connection $\gamma^i{}_{jk}(x,\dot x)$ can now be defined
in terms of the higher spin connections $\Ga{s}$,
\begin{align} \label{defgamma}
\gamma^i{}_{nk}(x,\dot x) \equiv \sum_{s=1}^{\infty}
\Ga{s}^i{}_{n j_2...j_s k} \dot x^{j_2} \cdots 
 \dot x^{j_s} \,
\end{align}
which transforms as required:
\begin{align}
\delta \gamma^i{}_{nk}(x,\dot x) &= \ve^i{}_m \gamma^m{}_{n k} - 2
\ve^m{}_{(n} \gamma^i{}_{k)m} -\ve^i{}_{,n k} \,.
\end{align}
The inhomogeneity $\ve^i{}_{,n k}$ follows from the variation $\delta
\Ga{1}$, while the terms with $s >1$ on the r.h.s.\ of
(\ref{defgamma}) transform as (use $\delta \dot x^i= \dot \ve^i =
\ve^i{}_m \dot x^m$)
\begin{align}
\delta &( \Ga{s}^i{}_{nj_2...j_s k} \dot x^{j_2} \cdots \dot x^{j_s}) \\
&=(\ve^i{}_m \Ga{s}^m{}_{n j_2...j_s k}
 -2 \ve^m{}_{(n|} \Ga{s}^i{}_{j_2...j_s |k)m}) \dot x^{j_2} \cdots \dot x^{j_s}
\nonumber \,,
\end{align}
where only the indices $n$ and $k$ are symmetrized.  Here terms
involving the variations $\delta \dot x^i$ have cancelled $s-1$ terms
in the tensor transformation of $\Ga{s}$ ($s>1$).

Physically, Eq.~(\ref{defgamma}) means that a spacetime equipped with
higher spin connections is equivalent to a spacetime with a
velocity-dependent affine connection $\gamma^i{}_{jk}(x,\dot x)$. The
gravitational force on a test particle thus depends not only on the
location of the particle, but also on its velocity similar as in a
Finsler space. However, since $\gamma^i{}_{jk}$ is not derived from
any metric structure, this spacetime is more general than a Finsler
space. 
%
\subsection{Matter currents} \label{secadditional}
%

We have not yet discussed the matter currents associated with the
generalized connections. Here, we restrict ourselves to a few
comments. A thorough discussion of the matter currents is beyond the
scope of this paper.

Consider a general matter Lagrangian ${\cal L}={\cal L}(\Psi, d\Psi,
\vartheta^\alpha, d\vartheta^\alpha, \Ga{s}, d\Ga{s})$ which includes
a matter field $\Psi$, the coframe $\vartheta^\alpha$ and the
generalized connections $\Ga{s}^{\alpha}{}_{\beta_1...\beta_{s+1}}$
($s \geqslant 0$) as given by Eqns.~(\ref{coeff-1})--(\ref{redef}). We may
then define the $d$$-$$1$-form currents 
\begin{align}
\Sigma_{\alpha} &:= \frac{\delta \cal L}{\delta \vartheta^{\alpha}} \,,\\
\Delta^{(s)}_{\alpha}{}^{\beta_1...\beta_{s+1}} 
&:= \frac{\delta \cal L}{\delta \Ga{s}^{\alpha}{}_{\beta_1...\beta_{s+1}}}
\quad (s \geqslant 0)\,. 
\end{align} 
Here $\Sigma_{\alpha}$ is the canonical energy-momentum current and
$\Delta^{(s)}_{\alpha}{}^{\beta_1...\beta_{s+1}}$ denotes currents
which we will call {\em hypermomentum currents} of degree $s$.  The
currents $\Delta^{(s)}_{\alpha}{}^{\beta_1...\beta_{s+1}}$ generalize
the hypermomentum current $\Delta^{(0)}_{\alpha}{}^{\beta}$ known from
the Metric-Affine Theory of Gravity \cite{MAG}.  Hypermomentum is the
sum of the spin current
$\tau_{\alpha\beta}=\Delta^{(0)}_{[\alpha\beta]}$ and the shear and
dilation current $\Delta^{(0)}_{(\alpha\beta)}$.

The components of $\Sigma_{\alpha}=\Sigma_{k\alpha} dx^k$ and
$\Delta^{(s)}_{\alpha}{}^{\beta_1...\beta_{s+1}}=
\Delta^{(s)}_{k\alpha}{}^{\beta_1...\beta_{s+1}}dx^k$ may be used to
define the generators of the diffeomorphism algebra.  In fact,
integrating the components $\Sigma_{0\alpha}$ and
$\Delta^{(s)}_{0\alpha}{}^{\beta_1...\beta_{s+1}}$ over a
$d$$-$$1$-dimensional spacelike hypersurface, we recover (gauge
$\xi^i=x^i$)
\begin{align}
P_\alpha&=\int d^{d-1} x\, \Sigma_{0\alpha} \,,\\
F^{(s)}_{\alpha}{}^{\beta_1...\beta_{s+1}} &= \int d^{d-1} x\,
\Delta^{(s)}_{0\alpha}{}^{\beta_1...\beta_{s+1}} \,
\end{align}
which, by construction, satisfy the algebra (\ref{alg}).

Which are the matter fields carrying these currents?  Representations
of the Poincar\'e group carry only energy-momentum and spin.  In order to
have also sources for hypermomentum, we would have to construct field
equations for representations of the double covering of
$\overline{GL}(d, \RR)$ or the diffeomorphism group
$\overline{\textit{Diff}}\,(d,\mathbb{R})$. We briefly commented on
this in \cite{Kirsch}, Sec.~IVA, see also \cite{MAG}
and references therein.

%
\section{Conclusions} \label{sec5}
%
In this paper we discussed the higher-spin Goldstone fields
$\om{s}^i{}_{j_1...j_sk}$ of the spontaneous breaking of the group of
analytic diffeomorphisms and its relevance for gravity. It is quite a
challenge to construct a Higgs mechanism for the complete
diffeomorphism group. 

As a partial realization, we provided a Higgs mechanism for the
breaking of its linear subgroup down to the Lorentz group. Our model
predicts that gravity is modified at high energies by the exchange of
a massive spin-3 field. This field was identified as the totally
symmetric part of the nonmetricity field~$Q_{ijk}$.  In \cite{Kirsch}
we suggested the name ``triton'' 
for the
corresponding particle. The range of this additional spin-3 force is
of order of the Compton wavelength $\lambda_c = {h}/{m_Q c}$ and
appears to be extremely short-ranged. The mass $m_Q$ of nonmetricity
enters our model as a free parameter and has to be measured
experimentally. Under the assumption that our model is related to
hybrid inflation, we estimated $m_Q$ to be at $10^{15} - 10^{16}$~GeV.

Of course, one expects \cite{I&O} all higher spin fields $\om{s}$ to
become massive due to a similar Higgs effect. To gain insight into the
complexity of the Higgs effect, we therefore modeled also the
absorption process for $\om{2}^i{}_{j_1j_2k}$ adopting the
St\"uckelberg formalism. 

{}From the nonlinear realization discussion, it is clear that the
complete symmetry breaking of the diffeomorphism group should provide
a massless graviton and an infinite tower of massive higher spin
particles. This particle spectrum reminds to that of string
theory, but with the difference that here the fields acquire mass by a
Higgs mechanism. It would be exciting to find a constraint in a
generalization of our Higgs model to higher-spin fields which
constraints the corresponding particles to lie on Regge trajectories.

%
\section*{Acknowledgments}
%
We would like to thank N.~Arkani-Hamed, F.~W.~Hehl, A.~Krause,
J.~Nuyts and Ph.~Spindel for many useful discussions related to this
work.  N.B.~wants to thank N.~Arkani-Hamed for his invitation at the
High-Energy Physics Group of the Jefferson Laboratory.  I.K.~is
grateful to Ph.~Spindel for kind hospitality at the University of
Mons-Hainaut. 
N.B.~is a Postdoctoral Researcher of the Fonds National de la 
Recherche Scientifique (Belgium). 
The work of I.K.~was supported by a fellowship within
the Postdoc-Program of the German Research Society (DFG), grant
KI~1084/1.


\appendix

\section{The transformation behavior of the coset fields $\om{s}$} 
\label{app1}

In this appendix we compute the transformation behavior of the coset
fields $\om{s}^i{}_{j_1...j_sk}$ associated with the broken generators
$F_i^{(s)j_1...j_sk}$ of the diffeomorphism group. The computation is
analog to that in \cite{Kirsch}, App.~A, where it was performed for the
special case $s=1$.

For simplicity, we only consider fields with holonomic indices and 
restrict on the coset space $G/H=\textit{Diff}\,(d, \RR)/GL(d, \RR)$. For
the coset element $\sigma \in G/H$, the group elements $g \in G$ and $h \in
H$, we choose the parameterizations
\begin{align}
  \s(\xi,\omega) &= e^{i\xi\cdot P}e^{i\stackrel{(1)}{\omega}\cdot F^{(1)}}
  e^{i\stackrel{(2)}{\omega}\cdot F^{(2)}} \cdots \,, \\
  g(\e) &\approx 1 + i \e\cdot P + i\e\cdot F^{(0)} + i\e\cdot F^{(1)} + 
  \ldots\,,\\
  h(\alpha) &\approx 1 + i \alpha \cdot M \, , \quad \a=\a(\epsilon;\xi,
  \omega) \,,
\end{align}
where 
\begin{align}
  &\e\cdot P = \e^iP_i \,,\quad \e\cdot F^{(0)}=\e^i_{~j}F^{(0)j}_i\,, 
  \nonumber\\
  &\e\cdot F^{(1)}=\e^i_{~j_1j_2}F^{(1)j_1j_2}_i\,, \quad {etc}.  
\end{align}

In order to obtain the transformation behavior $\d \om{s}^i{}_{j_1...j_sk}$,
we substitute the above parametrizations into the nonlinear
transformation law for elements $\sigma$ of $G/H$ given by \cite{CCWZ,
Salam}
\begin{align} \label{A12}
g(\e)\s(\xi,\omega)=\s(\xi',\omega')h(\e,\xi,\omega)\, .
\end{align}
Solving for $h(\e,\xi,\omega)$, we get
\begin{widetext}
\begin{align}
1+ i \alpha \cdot F^{(0)} = & 
\cdots
e^{-i\stackrel{(2)}{\omega}\cdot F^{(2)}}
e^{-i\stackrel{(1)}{\omega}\cdot F^{(1)}}
(1 + i\stackrel{(0)}{\ve}\cdot F^{(0)} 
+ i\stackrel{(1)}{\ve}\cdot F^{(1)} + \ldots)
e^{i\stackrel{(1)}{\omega}\cdot F^{(1)}}
e^{i\stackrel{(2)}{\omega}\cdot F^{(2)}}\cdots  \label{hsgs} \\
+ & \sum_{n=1}^\infty \cdots
e^{-i\stackrel{(n+1)}{\omega}\cdot F^{(n+1)}}
e^{-i\stackrel{(n)}{\omega}\cdot F^{(n)}}
\Big(i \sum_{i_n=0}^\infty \frac{(-1)^{i_n+1}}{(i_n+1)!}
(\stackrel{(n)}{\omega})^{i_n} [\delta\! \stackrel{(n)}{\omega}] \cdot
F^{(n i_n+n)}\Big)
e^{i\stackrel{(n)}{\omega}\cdot F^{(n)}}
e^{i\stackrel{(n+1)}{\omega}\cdot F^{(n+1)}}\cdots
\,,\nonumber
\end{align} 
\end{widetext}
where
\begin{align}
&\stackrel{(n)}{\ve}\cdot F^{(n)} =\frac{1}{(n+1)!} 
\frac{\partial^{n+1}\ve^i(\xi)}{\partial\xi^{j_1}\ldots \partial\xi^{j_n}
\partial\xi^k} F^{(n)j_1\ldots j_n k}_i\,,\nonumber \\ 
& \ve^i(\xi)  \equiv
\e^i + \e^{i}_j\xi^j + \e^{i}_{j_1j_2}\xi^{j_1}\xi^{j_2} + \ldots
= \d \xi^i \,.
\end{align}
Note that we have already performed the multiplication with 
$e^{\pm i \xi \cdot P}$. As shown in detail in \cite{Kirsch}, App.~A1, 
this promotes the parameters of $g(\epsilon)$ to space-time dependent fields:
$\epsilon \rightarrow \varepsilon(\xi)$. 

For the computation of (\ref{hsgs}), it turns out to be convenient to
introduce the following bracket notation: For any two tensors
$T^{(n)}\leadsto T^i_{~j_1\ldots j_n k}$ and $U^{(q)}\leadsto
U^i_{~j_1\ldots j_q k}$ of type $(1,n+1)$ and $(1,q+1)$, completely
symmetric in their covariant indices, we have the following bracket
which gives a tensor of type $(1,n+q+1)$, completely symmetric in its
covariant indices as well
\begin{align}
[~,~]\quad : \quad &(T^{(n)},U^{(q)}) \longrightarrow 
[T^{(n)},U^{(q)}]^{(n+q)} \,,\nonumber\\
&{[T^{(n)},U^{(q)}]}^i_{~j_1\ldots j_{n+q}k}=\nonumber\\
&\hspace{0.5cm}(n+1)\,T^i_{~l(j_1\ldots j_{n}}
U^l_{~j_{n+1}\ldots j_{n+q}k)} \nonumber\\
&\hspace{0.5cm}- (q+1)\,U^i_{~l(j_1\ldots j_{q}}
T^l_{~j_{q+1}\ldots j_{n+q}k)}\,.
\label{bra}
\end{align}
If we further define the notation 
\begin{align}
\stackrel{(n)}{\omega}[\stackrel{(p)}{\ve}]&:=[\stackrel{(p)}{\ve},\stackrel{(n)}{\omega}]\,,
\nonumber \\
(\stackrel{(n)}{\omega})^2[\stackrel{(p)}{\ve}]&:=
[[\stackrel{(p)}{\ve},\stackrel{(n)}{\omega}],\stackrel{(n)}{\omega}]\,,\nonumber \\
(\stackrel{(n)}{\omega})^3[\stackrel{(p)}{\ve}]&:=
[[[\stackrel{(p)}{\ve},\stackrel{(n)}{\omega}],\stackrel{(n)}{\omega}],\stackrel{(n)}{\omega}]
\,,\nonumber \\
&\vdots
\nonumber 
\end{align}
and
\begin{align}
\stackrel{(n_1)}{\omega} \cdots \stackrel{(n_s)}{\omega}[\stackrel{(p)}{\ve}]&:=
[...[\stackrel{(p)}{\ve},\stackrel{(n_s)}{\omega}],\stackrel{(n_{s-1})}{\omega}],
...],\stackrel{(n_1)}{\omega}] \,,
\end{align}
then, for example,
\begin{align}
e^{-i\stackrel{(r)}{\omega}\cdot F^{(r)}}
(1 + i\stackrel{(0)}{\ve}\cdot F^{(0)} 
+ i\stackrel{(1)}{\ve}\cdot F^{(1)} + \ldots)
e^{i\stackrel{(r)}{\omega}\cdot F^{(r)}} 
\nonumber \\
= 1 + i \sum_{s=1}^{\infty}
\Big( \sum_{k=0}^{s-1} \frac{1}{k!}
(\stackrel{(r)}{\omega})^k [\stackrel{(s-k-1)}{\ve}]\Big)\cdot F^{(s-1)} \,.
\end{align}
Here we used the Baker-Campbell-Hausdorff formula in the form
\begin{align}
e^{-B} A e^{B} = A + [A,B] + \frac1{2!} [[A,B],B] + ... 
\end{align}
for two operators $A$ and $B$.

We have now an algorithm to write down $\d\! \stackrel{(s)}{\omega}$ in a closed 
form. Comparing successively the coefficients of $F^{(1)}$, $F^{(2)}$,
$F^{(3)}$, etc.\ in Eq.~(\ref{hsgs}), we obtain 
\begin{align}
\d\! \stackrel{(1)}{\omega} &=\stackrel{(1)}{\ve} + 
\stackrel{(1)}{\omega}[\stackrel{(0)}{\ve}] \,,\label{dom1}\\
\d\! \stackrel{(2)}{\omega} &=\stackrel{(2)}{\ve} + 
\stackrel{(2)}{\omega}[\stackrel{(0)}{\ve}] + \frac{1}{2}
\stackrel{(1)}{\omega}[\stackrel{(1)}{\ve}] \,, \label{dom2}\\
\d\! \stackrel{(3)}{\omega} &= \stackrel{(3)}{\ve} + 
\stackrel{(3)}{\omega}[\stackrel{(0)}{\ve}] + 
\stackrel{(1)}{\omega}[\stackrel{(2)}{\ve}] - \frac{1}{3!}
\stackrel{(1)}{\omega}\stackrel{(1)}{\omega} [\stackrel{(1)}{\ve}] \,, \\
etc.\ \nonumber
\end{align}
For general $\d\!\! \stackrel{(s)}{\omega}$, we therefore get
\begin{align}
\d\! \stackrel{(s)}{\omega}  = \stackrel{(s)}{\ve} + 
\stackrel{(s)}{\omega}[\stackrel{(0)}{\ve}] + ... \,,
\end{align}
which is identical to Eq.~(\ref{transformation}).
%
%
\section{The total nonlinear connection} \label{appB}
%
In this appendix we give a compact expression for the total nonlinear
connection $\Gamma=\sigma^{-1} d\sigma$. The coset element $\sigma \in
\textit{Diff}\,(d,\mathbb{R})/SO(1,d-1)$ will be parameterized as in
Eq.~(\ref{cosetelement}). After a short computation, we get
\begin{align}
&\Gamma = \\
 &\sum^\infty_{n = -1 } 
\prod^{\infty}_{m=n+1} e^{-i \om{m} \cdot F^{(m)}} 
(e^{-i\om{n}\cdot F^{(n)}} d e^{i\om{n}\cdot F^{(n)}}) 
e^{i \om{m} \cdot  F^{(m)}} ,  \nonumber
\end{align}
where
\begin{align} 
e^{-i\om{n}\cdot F^{(n)}} d e^{i\om{n}\cdot F^{(n)}}
=  \sum_{i_n=0}^{\infty} \frac{i(\om{n})^{i_n} [d\om{n}]}{(i_n+1)!}  \cdot
F^{(ni_n+n)} \nonumber\,.
\end{align}
Here we defined \mbox{$\om{-1} \equiv \xi$} and \mbox{$\om{0} \equiv
h$}, where $h$ is the shear coset parameter corresponding to
$GL(d,\RR)/SO(1,d-1)$. It is understood that the exponentials have to
be written in ascending (descending) order on the right (left) of the
central factor $e^{-...} d e^{...}$.

Using the bracket notation of App.~\ref{app1}, we find for the
1-forms $\Gamma^i{}_{j_1..j_{s}}=(\Gamma
\vert_{F^{(s-1)}})^i{}_{j_1..j_{s}}$ the compact expression
\begin{align}
\Gamma \vert_{F^{(s-1)}}=& \!\sum^{s-1}_{n = -1} 
\sum_{i_{n},\ldots,i_s=0 }^s 
(\om{s})^{i_{s}} \cdots (\om{n})^{i_{n}}  [d\om{n}] \nonumber\\ 
&\times \frac{\delta(i_{n},\ldots,i_s, n)}{(i_n+1)!i_{n+1}!\ldots i_s!}
\delta_{i_{-1},0}\,,  \label{SchoutenGamma}
\end{align}
where
\begin{align}
\delta(i_n, i_{n+1},\ldots,i_s, n) = 1 \,,
\end{align}
if $n+n i_n+(n+1)i_{n+1}+\ldots+ s i_s=s-1$, zero otherwise. 

To linear order this can be expanded as
\begin{align}
\Gamma \vert_{F^{(s-1)}} =  d \om{s-1} + \om{s} [d \xi] + 
{\cal O}(\omega^2) \,,
\end{align}  
where the first two terms correspond to $n=s-1, i_n=0$ and
$n=-1, i_1=...=i_{s-1}=0, i_s=1$, respectively. 

The first five coefficients are
\begin{align}
\Gamma \vert_{F^{(-1)}} &=\vartheta \equiv e^h [d\xi] = 
(1+h+{\frac{1}{2}}h^2 + \ldots)[d\xi] \,, \label{GammaF-1}\\ 
\Gamma \vert_{F^{(0)}} &=  e^{-1} d e+ \om{1} [\vartheta]  \,,\\
\Gamma \vert_{F^{(1)}} &=  d \om{1} + \big( \om{2}
+ \frac{1}{2!} (\om{1})^2 \big)[\vartheta] + \om{1}[e^{-1} d e] \,, 
\label{GammaF1}\\
\Gamma \vert_{F^{(2)}} &=  d \om{2} + \big(\om{3}  + 
\om{2}\om{1} +\frac{1}{3!} (\om{1})^3  \big) [\vartheta] \\
&\hspace{.5cm}+ \big(\om{2}+{\frac{1}{2!}}(\om{1})^2\big)[e^{-1} d e]
+ \frac{1}{2!} \om{1}[d\om{1}] \,, \nonumber\\
\Gamma \vert_{F^{(3)}} &=  d \om{3} + \big(\om{4}  + 
\om{3}\om{1} + \frac{1}{2!} (\om{2})^2  \nonumber\\
&\hspace{.5cm}+ \frac{1}{4!} (\om{1})^4 \big) [\vartheta] \nonumber\\
&\hspace{.5cm}+ \big(\om{3}  + 
\om{2}\om{1} +\frac{1}{3!} (\om{1})^3  \big)[e^{-1} d e] \nonumber\\
&\hspace{.5cm}+ \big( \om{2}+ \frac{1}{3!} (\om{1})^2 \big) [d\om{1}] \,,
\end{align}
with
\begin{align}
e^{-1} d e = \sum_{i_0=0}^{\infty} {\frac{1}{(i_0+1)!}} h^{i_0} [dh] \,.
\end{align}
If we apply the rule for the bracket in Eq.~(\ref{bra}), we find 
Eqs.\ (\ref{coeff-1}) -- (\ref{redef}).

The transformation law for the 1-forms
$\Gamma^i{}_{j_1..j_{s}}=(\Gamma \vert_{F^{(s-1)}})^i{}_{j_1..j_{s}k}
d\xi^k$ follows from those for $\om{s}$ given in
App.~\ref{app1}. Since $\Gamma_{G/H}$ transforms as a tensor, we
expect
\begin{align}
\delta \Gamma\vert_{F^{(s-1)}} &= \big(\Gamma\vert_{F^{(s-1)}} \big) 
[\stackrel{(0)}{\ve}]=[\stackrel{(0)}{\ve}, \Gamma\vert_{F^{(s-1)}}] 
\end{align}
for $s \geqslant 2$. We explicitly checked this for
$\Gamma\vert_{F^{(1)}}$ using Eqs.\ (\ref{GammaF1}), (\ref{dom1}) and
(\ref{dom2}). 
%
%
\section{Decomposition of higher spin connections} \label{app3}
%
Upon lowering the upper index $i$, the higher spin connection
$\omega^{i}{}_{j_1...j_{s}k}$ can be decomposed under $GL(d, \RR)$
into a totally symmetric part corresponding to the Young tableau
$[s+2,0]$ and a part corresponding to $[s+1,1]$:
\begin{align}
 \hspace{2cm} &GL(d,\, \RR) \hspace{1.5cm}  {\rm dimension} \nonumber\\
\raisebox{0.1cm}{$\om{s}_{(ij_1...j_{s}k)}$} \quad 
&\begin{picture}(80,10)
      \put(0,0) {\line(1,0){60}}              %
      \put(0,10) {\line(1,0){60}}
      \multiput(0,0)(10,0){2} {\line(0,1){10}} 
      \multiput(50,0)(10,0){2} {\line(0,1){10}}  
      \put(21,1) {$\cdots$}   
     \end{picture}   \frac{(d+s+1)!}{(d-1)!(s+2)!} \,, \nonumber \\
\raisebox{0.1cm}{$\om{s}_{[ij_1]...j_{s}k}$} \quad 
&\begin{picture}(80,10)
      \put(0,0) {\line(1,0){50}}              %
      \put(0,10) {\line(1,0){50}}
      \multiput(0,0)(10,0){2} {\line(0,1){10}} 
      \multiput(40,0)(10,0){2} {\line(0,1){10}}  
      \put(21,1) {$\cdots$} 
       \put(0,-10) {\line(1,0){10}}              
      \put(0,-10) {\line(1,0){10}}
       \multiput(0,-10)(10,0){2} {\line(0,1){10}} 
     \end{picture} \frac{(d+s)!(s+1)}{(d-2)!(s+2)!}  \,. \nonumber
\end{align}
In total, the higher spin connection $\om{s}$ has 
\begin{align} \label{offshell}
  d \begin{pmatrix} d+s \\s+1 \end{pmatrix}
\end{align} 
off-shell components. 

Let us consider the case in which $\om{s}$ is massless. Then, in order
to apply the Fronsdal description for these fields, we have to split
$\om{s}$ into double-traceless fields. For instance,
$\om{s}_{(ij_1...j_{s}k)}$ is equivalent to the sum of
double-traceless fields $\omhathat{s}_{(ij_1...j_{s}k)}$, $
\omhathat{s}_{(ij_1...j_{s-4}k)}$, $
\omhathat{s}_{(ij_1...j_{s-8}k)}$, etc.

The number of on-shell degrees of freedom are given by the same Young
diagram, now labeling an $O(d-2)$ irreducible representation.  In $d =
4$, the fields $\omhathat{s}_{(ij_1...j_{s-n}k)}$ ($n=0,4,8,...$),
have spin $s-n+2$ and $2$ on-shell degrees of freedom, while the
fields $\omhathat{s}_{[ij_1]...j_{s-n}k}$ ($n=0,4,8,...$) are {\em
non-dynamical}. The ``hook'' representations have vanishing on-shell
degrees of freedom, since the dimension of the same Young tableau
under the little group $O(2)$ is zero.

\end{document}